\documentclass[12pt]{article}
\usepackage[margin=1.125in]{geometry}

\pdfoutput=1

\usepackage{thesis}
\begin{document}
 
\begin{center}~\\
{\it Implications of Superrotations}
~\\~\\~\\~\\
A dissertation presented\\
by\\
Sabrina Pasterski\\
to\\
The Department of Physics\\
in partial fulfillment of the requirements\\
for the degree of\\
Doctor of Philosophy\\
in the subject of\\
Physics\\
~\\~\\~\\~\\~\\
Harvard University\\
Cambridge, Massachusetts\\
April 2019\\
\end{center}
~\\
\pagenumbering{roman}
\setcounter{page}{1}
\thispagestyle{empty}  \addcontentsline{toc}{subsection}{Title Page}

\newpage

~\\~\\~\\~\\~\\~\\~\\~\\~\\

\addcontentsline{toc}{subsection}{Copyright}
\indent\indent\indent\indent\indent\indent\indent\indent\indent{\small \textcopyright}~2019 Sabrina Pasterski\\
\indent\indent\indent\indent\indent\indent\indent\indent\indent All rights reserved.

~\\
\pagestyle{plain}\setcounter{page}{2}\thispagestyle{empty}
\pagebreak

\pagestyle{plain}
\pagenumbering{roman}
\noindent Dissertation Advisor: Andrew Strominger\indent\indent\indent\indent\indent\indent\indent\indent~~~~~~ Sabrina Pasterski
~\\
\begin{center} Implications of Superrotations\\
~\\
{\bf Abstract}
 \end{center}

A framework of connections between asymptotic symmetries, soft theorems, and memory effects has recently shed light on a universal structure associated with infrared physics.  Here, we show how this pattern has been used to fill in missing elements.  After the necessary groundwork, we begin by proving a Ward identity for superrotations using the subleading soft graviton theorem, thereby demonstrating a semiclassical Virasoro symmetry for scattering in quantum gravity.  Next, we show there exists a new spin memory effect associated with this symmetry, explain more generally how the connections between the vertices of the infrared triangle predicted this, and describe what other examples and variations have been unveiled.  Taking to heart this newly motivated Virasoro symmetry, we review how the soft theorem has been recast as a Virasoro Ward identity for a putative two dimensional conformal field theory.  This derivation relies upon a map from plane wave scattering states to a conformal primary basis, which we then construct.  We provide examples of familiar scattering amplitudes recast in this basis and discuss the somewhat exotic nature of the putative CFT$_2$.  We conclude by describing ongoing efforts to tame some of these features and what this change of basis  in turn has taught us about the infrared limit which began  our story.

\addcontentsline{toc}{subsection}{Abstract}

\setcounter{page}{3}

\pagebreak 
\begin{center} 
{\vspace{-6.7em}\bf Table of Contents}
 \end{center}\vspace{-5.7em}

{\fontsize{12}{12}
\def\contentsname{\empty}
 \addcontentsline{toc}{subsection}{Table of Contents}
\tableofcontents}

\pagebreak
~\\~\\~\\
\begin{center} 
{\bf Acknowledgements}
 \end{center}
 
 I am indebted to the guidance, instruction, and insights of my advisor, committee, collaborators, and colleagues at the Center for the Fundamental Laws of Nature.  This thesis was completed with generous support from the National Science Foundation through a Graduate Research Fellowship under grant DGE-1144152 and the Hertz Foundation through a Harold and Ruth Newman Fellowship.
 \addcontentsline{toc}{subsection}{Acknowledgements}
 \pagebreak

 \indent \indent \indent \indent\indent\indent\indent\indent\indent\indent\indent\indent\indent\indent\indent\indent\indent\indent\indent\indent\indent\indent{\it To the future...}
  \addcontentsline{toc}{subsection}{Dedication}
 \pagebreak

\pagenumbering{arabic}

\section{Introduction}

\indent\indent Our story starts with Strominger's suggestion that a series of separate studies from the sixties are secretly the same.  The relativists were systematizing what happens at long distances.  The quantum field theorists were worried about what was going on at low energies. And, a little later, someone remembered there was a physical observable attached to each of these things. Together they formed a triangle of traits universal enough to make new predictions.  When old gaps were filled in, new iterations popped up.

But, one copy of this triangle arrived with a twist.  It came prepared with an ingredient that would let us connect our story to an even bigger saga that has preoccupied our field for as long as some of us have any memories at all.  The idea that we can describe a theory about gravity without gravity, in a lower dimension -- an idea that allows us to geometrize the entanglement of a quantum theory or de-geometrize a theory of quantum gravity at will. 

However, the freedom to compute on whichever side is easiest is only earned by showing that both sides are equivalent.  While we are armed with a dictionary that lets us translate between the language of gravitational theories with a negative cosmological constant and field theories with  nice rescaling symmetries, the universe seems to be giving us the wrong sign. 

But if there are still some people who believe the Earth is flat, who can begrudge a theorist pretending the universe is flat when the  cosmological constant is so small anyway?  So we study scattering in asymptotically flat spacetimes; aim to find a holographic dual description; and know that despite what's still missing we've at least landed on some interesting statements about infrared physics.

This thesis studies asymptotic symmetries of asymptotically flat spacetimes with the hope that our efforts will teach us something about quantum gravity. But tacking a buzzword onto technical jargon is too evasive a tactic to serve as a justification.  At a basic level, we are doing what we can with what we know.  We know that the more symmetric a problem is, the more constrained its solutions are.  So, we look for more symmetries.  We think we can find more symmetries because we think we can relax assumptions made by others.  We believe this is a good idea because we have seen it work before.  Namely AdS/CFT.  Indeed, failing to attach the name `flat space holography' to our efforts {\it would} make our endeavors sound significantly less sweet.  Whether one prefers a montage of high hopes to a Montague, or seeks to question the {\underline{q}}uality of our putative {\underline{d}}uality, one letter makes a difference: and we must be start to be precise.   
The games are over, let the fun begin.

 \begin{figure}[bth!]
\centering
\includegraphics[clip, trim=3.5cm 13.5cm 3.5cm 3.5cm, width=.8\textwidth]{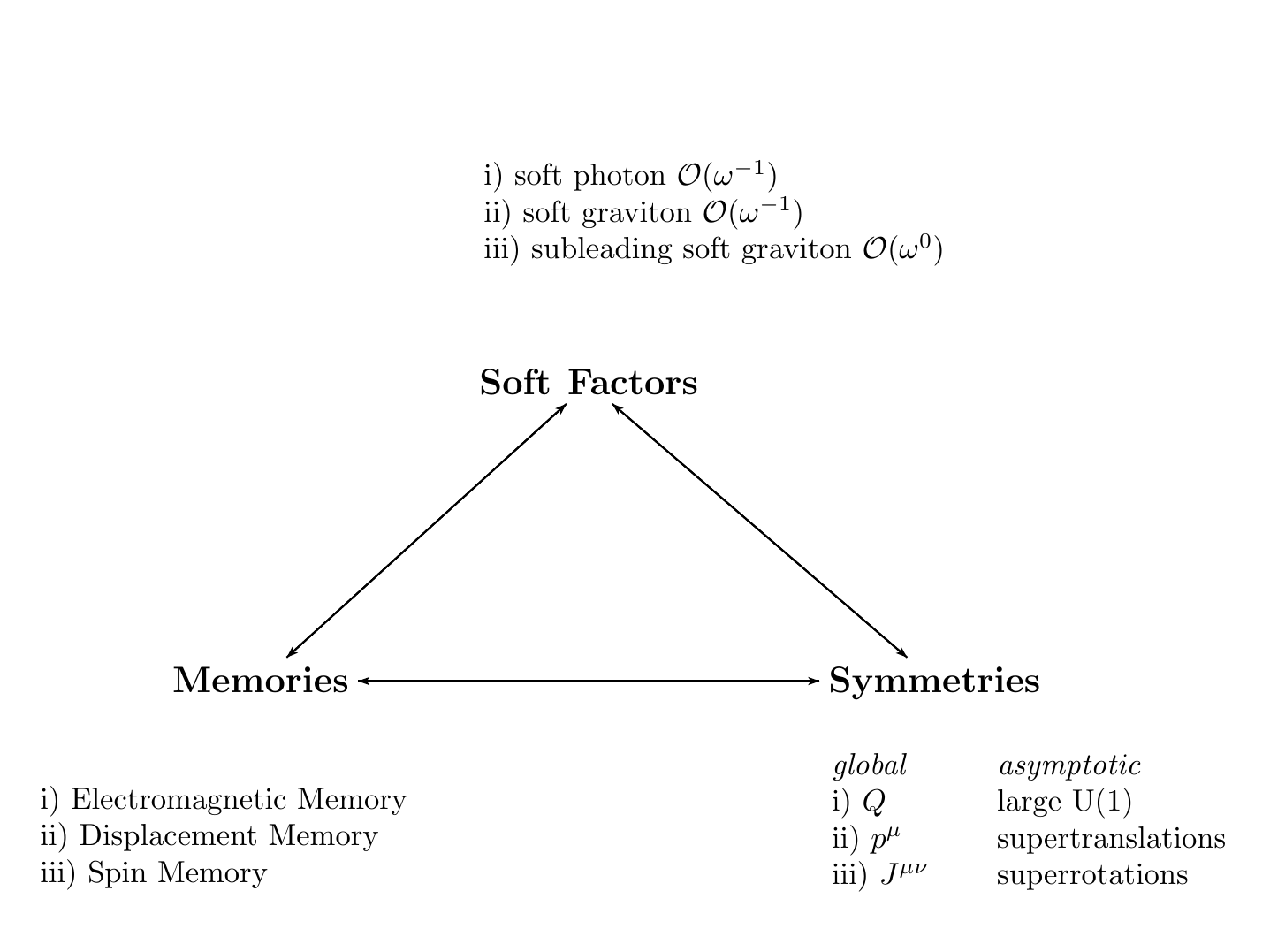}
 \caption{The IR Triangle.  There exists a series of universal connections within infrared physics, which we are tasked to explain, exploit, and expand upon herein. 
 }\label{fig:IRtriangle}
\end{figure}
\newpage

The motivation for this thesis is two-fold.  First, there is a set of intimate connections between diverse aspects of infrared physics that has allowed us to make predictions about new symmetries of asymptotically flat spacetimes.  Second, these new predictions appear to be important building blocks for a flat space rendition of holography.  Consider the infrared triangle of Figure~\ref{fig:IRtriangle}.  We will get to the backstory of each vertex and see their connections in due course, but let us pause to consider what these objects are and why they are related.  At the lower right we have `symmetries.'  These give us conservation laws and, in the setting of gauge theories, act non-trivially on our phase space when the generators don't fall off at the boundary of our spacetime.  However, we will soon see that the boundary of asymptotically flat spacetimes is null, and quantities which are charged under these symmetries can thus escape along it.  There are some very low energy `memory' modes which correspond to net changes that occur along this boundary and they show up as IR divergences in quantum field theory amplitudes with universal residues called `soft theorems'.  One interesting application of this triangle has been a re-interpretation of IR divergence issues in terms of charge non-conservation for asymptotic symmetries.  So our understanding of the connections in the triangle has paid off in this respect.

For the purpose of this thesis, we are most concerned with the fact that this triangle predicts more copies of itself.  The three examples explicitly listed in Figure~\ref{fig:IRtriangle} are only a taste and the story extends even beyond gauge theories.  The listed  iterations are the ones we will talk about here because the first two -- electromagnetic and leading gravitational -- were well established at a time when the third was essentially brand new.  Our understanding of the superrotations we care about here was built around this framework.

But once we had superrotations, an extension of the asymptotic symmetry group of asymptotically flat spacetimes to include local conformal transformations of the night sky, we also had a key ingredient in motivating a 2D holographic dual to flat space scattering -- we had a second framework to build, namely a map from 4D to 2D.  The first things we knew how to map were the soft modes to celestial sphere currents, and the 4D Ward identities to 2D Ward identities for them.  The latter portion of this thesis shows how to extend this map beyond the soft limit.

We begin with some groundwork, defining asymptotically flat spacetimes in section~\ref{sec:asym}, in particular their causal structure in~\ref{sec:nullinf} and the universality of their behavior at large distances in~\ref{sec:ashtekar}, as well as what this implies for a perturbative expansion of the metric in~\ref{sec:bondi}, restricting to four spacetime dimensions.  We then describe how to apply this to the scattering problem in section~\ref{sec:scat}, where we will need tools from quantum field theory regarding soft limits of gauge boson insertions in~\ref{sec:soft}.  We lay out some conventions for handling mode expansions in~\ref{sec:mode}, before getting to the core of this thesis in section~\ref{sec:superR}.  Here, we present the results of \cite{1406.3312KLPS,1502.06120PSZ,Pasterski:2015zua,1902.01840HMP}.  We define different aspects of the infrared triangle of Figure~\ref{fig:IRtriangle}.  We use the soft theorem from~\ref{sec:soft} to prove a Ward identity for superrotations in~\ref{sec:Ward}.  We describe how the soft modes connect to physical observables in~\ref{sec:mem}, touching on the three examples of memory effects listed in Figure~\ref{fig:IRtriangle} in~\ref{sec:emem},~\ref{sec:gdis}, and~\ref{sec:spinmem} before making some comments on the connection between memory effects and asymptotic symmetries in section~\ref{sec:vactr}.  We then present a key step in going from superrotations to their implications in section~\ref{sec:stress}, when we describe how the 4D subleading soft graviton mode appears to act as a 2D stress tensor.  We then transition to describing the tools we've developed in~\cite{Pasterski:2016qvg,Pasterski:2017ylz, Pasterski:2017kqt} to try to see where these implications take us in section~\ref{sec:cb}.  We describe the scattering basis preferred by superrotations in~\ref{sec:pb}, outlining our conventions in~\ref{sec:conventions} and find the explicit wavefunctions and weight spectrum in~\ref{sec:pbasis}, before getting to some examples of what familiar amplitudes actually look like in this basis in~\ref{sec:examples}.  Finally we assess the current state of affairs and aspirations in~\ref{sec:unitary}.

\newpage 
\section{Asymptotically Flat Spacetimes}\label{sec:asym}
In this section we define what we mean by asymptotically flat spacetimes and asymptotic symmetries thereof.  The point is to formalize the notion of spacetimes for which the cosmological constant is zero and the matter sources are localized.  We are considering solutions to Einstein's equations
\be
R_{\mu\nu}-\frac{1}{2}g_{\mu\nu} R=8\pi G T_{\mu\nu}
\ee
where we expect the matter stress tensor to `fall-off' at a certain rate as one moves far away, and the metric in this asymptotic region to approach that of flat spacetime.  This raises the questions of what it means to `go to infinity,' how the metric should approach flatness, and moreover what is the minimal set of assumptions needed to make such statements.  

For most of this thesis we will be taking a very explicit coordinate-based approach, initiated by Bondi, van der Burg, Metzner, and Sachs (BMS)~\cite{doi:10.1098/rspa.1962.0161,doi:10.1098/rspa.1962.0206}, of writing down an expansion of the metric in powers of a radial coordinate and looking at the class of diffeomorphisms that preserve these falloffs to identify the asymptotic symmetry group.  This expansion will be around a region of spacetime that captures the `radiation zone,' reached by null geodesics at infinite affine parameter.  

In order to appreciate and interpret this expansion, we thus need to answer our first question above: what it means to `go to infinity.'  The next subsection will introduce the conformal compactification of Minkowski space, so that we can familiarize ourselves with the different notions of infinity that exist for flat and asymptotically flat spacetimes.  We will then take a brief moment in section~(\ref{sec:ashtekar}) to quote the formal definition of asymptotically flat spacetimes in terms of conformal compactifications to give a better picture of how the other questions raised above have been answered~\cite{Ash1,Ash2,Ash3}, before returning to the Bondi expansion for the computations needed to demonstrate our new results.

\subsection{Null Infinity}\label{sec:nullinf}
\begin{figure}[th]
\centering
\includegraphics[width=.8\textwidth]{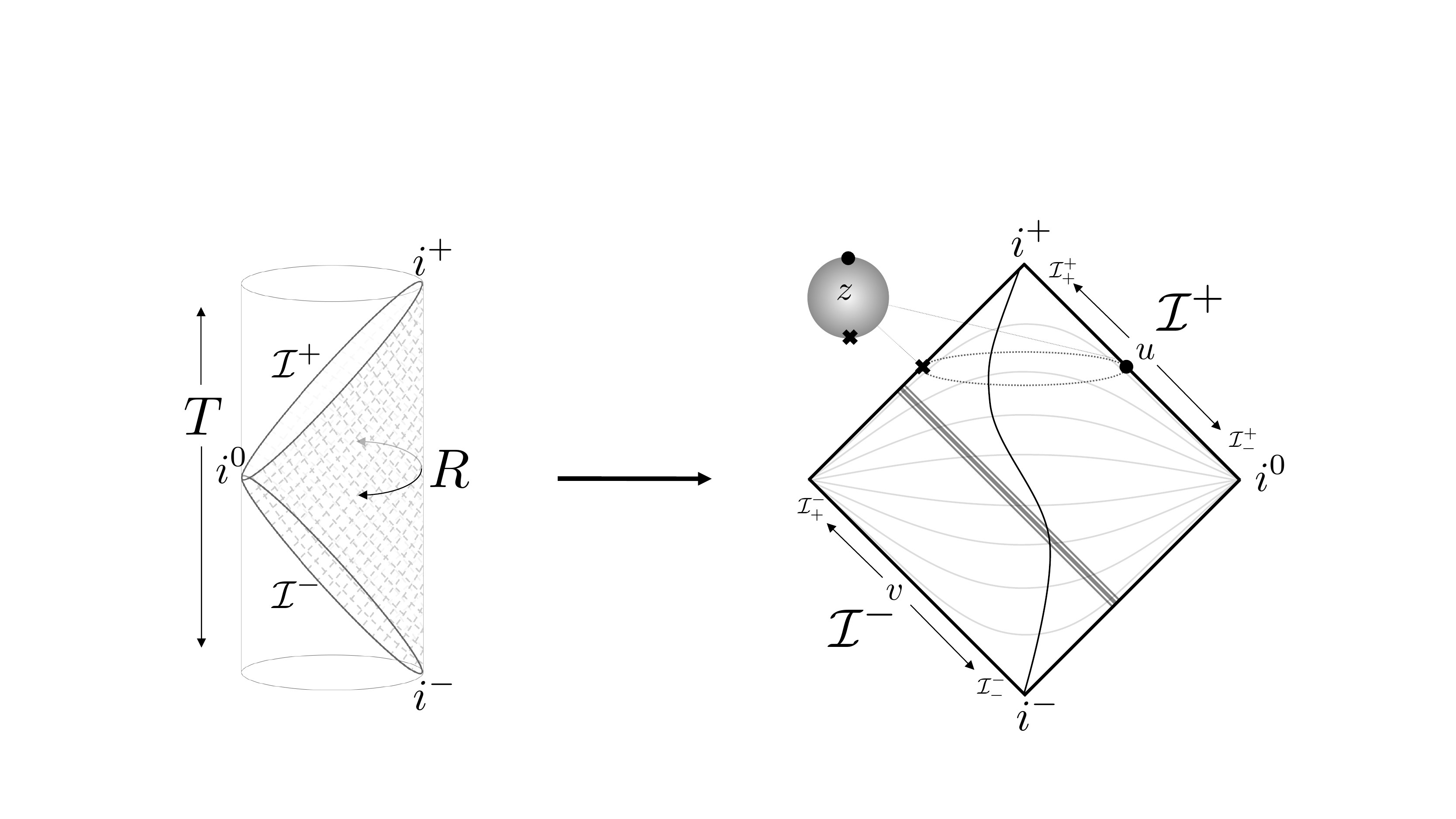}
 \caption{Penrose diagram for Minkowski space, represented as a patch of the Einstein static universe and unwrapped with antipodal points shown.  Massless trajectories travel at $45^\circ$ angles and enter and exit at $\mathcal{I}^\pm$.  Geodesics for massive particles enter at $i^-$ and exit at $i^+$.
 }\label{fig:penrose}
\end{figure}

In this subsection we construct the Penrose diagram of Minkowski space so that we can understand its causal structure~\cite{Penrose}.  The main takeaway will be an introduction to the notion of null infinity.  En route we will also be introducing coordinate conventions and notations that will be used throughout.  Let us start with the flat metric on Minkowski space in spherical coordinates
\be
ds^2=-dt^2+dr^2+2r^2 \gamma_{z\bz} dzd\bz
\ee
where we use projective coordinates for the $S^2$ factor, which we will often refer to as the `celestial sphere' ($\mathcal{CS}^2$).  The unit round metric is given by
\be\label{eq:spheremetric}
\gamma_{z\bz}=\frac{2}{(1+z\bz)^2}.
\ee
In terms of usual polar coordinates, $z=e^{i\phi}\tan\frac{\theta}{2}$. We can introduce the retarded and advanced times
 \be
 u=t-r,~~v=t+r
 \ee
 to label radially outgoing, respectively incoming, null geodesics -- for an outgoing radial geodesic $u$ is fixed and the value of $u$ labels when it was emitted from the origin.  For future reference, the Cartesian coordinates are now:
  \bea\label{flt}
x^0&=&u+r=v-r,  \cr
  x^1+ix^2&=&{2rz \over 1+z\bz},\cr
  x^3&=& {r(1-z\bz) \over 1+z\bz}.         
  \eea
We now perform the conformal compactification (see also chapter 11 of~\cite{Wald} and~\cite{Compere}).  Here $u,v\in(-\infty,\infty)$, however one can transform to rescaled coordinates 
  \be
 u=L\tan U,~~v=L\tan V
 \ee
 where the finite range $U,V\in (-\frac{\pi}{2},\frac{\pi}{2})$ covers our original manifold.  
Trading $(U,V)$ double null coordinates for a new $(T,R)$ so that 
\be
T=U+V,~~R=V-U
 \ee 
 we find that the Minkowski metric in these coordinates is the Lorentz metric on $S^3\times \mathbb{R}$ times a conformal factor:
 \be
ds^2=\frac{L^2}{4\cos^2(\frac{R-T}{2})\cos^2(\frac{R+T}{2})}(-dT^2+dR^2+2\sin^2 R \gamma_{z\bz} dzd\bz)=\Omega^{-2} d\tilde{s}^2
 \ee
where our coordinates cover the patch shown in Figure~\ref{fig:penrose}.  Because the conformal factor
\be
\Omega^{-2}=\frac{L^2}{4\cos^2(\frac{R-T}{2})\cos^2(\frac{R+T}{2})}
\ee
 is positive, the rescaled metric $d\tilde{s}^2$ will preserve the causal structure so curves that are timelike, null, or spacelike respectively remain as such with respect to this rescaled metric.  In particular, we can attach a boundary in this compactification and understand the different notions of `infinity' relevant to Minkowski space and, more generally, the asymptotically flat spacetimes we will be interested in:
 \begin{itemize}
 \item Massive particles following time-like trajectories enter at {\it past timelike infinity}, denoted as $i^-$ and parameterized by $(R,T)=(0,-\pi)$; and exit at {\it future timelike infinity}, denoted as $i^+$ and parameterized by $(R,T)=(0,\pi)$.
 \item Massless particles enter along {\it past null infinity}, denoted as $\mathcal{I}^-$ and parameterized by $U=-\frac{\pi}{2}, V\in(-\frac{\pi}{2},\frac{\pi}{2})$; and exit along {\it future null infinity}, denoted as $\mathcal{I}^+$ and parameterized by $V=\frac{\pi}{2}, U\in(-\frac{\pi}{2},\frac{\pi}{2})$.
 \item Moving along any spacelike trajectory eventually lands one on {\it spacelike infinity}, denoted as $i^0$ and parameterized by $(R,T)=(\pi,0)$.
 \end{itemize}
 In the following we will denote the future and past boundaries of future null infinity by $\mathcal{I}^+_\pm$  with the subscript referring to the sign as $u\rightarrow \pm \infty$.  Similarly, the boundaries of past null infinity as $v\rightarrow\pm\infty$ are denoted $\mathcal{I}^-_\pm$.  The fact that fields at null infinity can have non-trivial angular dependence in these limits is one harbinger of the fact that the above mapping is singular at $i^{\pm,0}$, which are each mapped to points in Figure~\ref{fig:penrose}.  Timelike and spatial infinity can be appropriately resolved, however the limits at $\mathcal{I^\pm_\pm}$ will be all that we need here.

\subsection{Coordinate-Free Definition}\label{sec:ashtekar}

With this example of the conformal compactification of Minkowski space under our belt and having distinguished the different notions of infinity that appear, we take a moment to provide the formal definition of asymptotic flatness, as codified in the works of Ashtekar from the 80s~\cite{Ash1,Ash2,Ash3} (see~\cite{1409.1800ash,1808.07093ash} for recent reviews).  One of the takeaways is that conformal compactifications -- often introduced to examine global features of the causal structure, and in the last subsection to define null infinity as a component of the boundary of compactified Minkowski space --  can serve as a starting point in an analysis of asymptotically flat spacetimes that avoids relying upon a particular coordinate expansion.  In essence, the structure of this boundary is what is universal when we move from flat spacetime to any asymptotically flat solution.  We now quote the definition given in~\cite{1409.1800ash}. 

\noindent{\bf Definition 2.1} A spacetime $(\hat{M},\hat{g}_{ab})$ will be considered {\it asymptotically flat at null infinity} if it is diffeomorphic to the interior $M\backslash \mathcal{I}$ of a conformal completion $(M,g_{ab})$ with boundary $\mathcal{I}$ such that:
\begin{itemize}
\item There exists a smooth conformal factor $\Omega$ such that in the interior $g_{ab}=\Omega^2\hat{g}_{ab}$, and on $\mathcal{I}$ we have $\Omega=0$ with $n_a\equiv\nabla_a\Omega$ nowhere vanishing.
\item $\hat{g}_{ab}$ is a solution to Einstein's equations with zero cosmological constant and a matter stress tensor $\hat{T}_{ab}$ such that $\Omega^{-2}\hat{T}_{ab}$ has a smooth limit to $\mathcal{I}$.
\item $\mathcal{I}$ has topology $S^2\times\mathbb{R}$.
\end{itemize}
This definition is restricted to an analysis near one of either $\mathcal{I}^+$ or $\mathcal{I}^-$.  $\Omega$ plays the role of $\frac{1}{r}$ as one approaches the boundary from null directions. (We inverted our convention for $\Omega$ as compared to~\cite{Wald} in the previous subsection so that one can check this is indeed how $\Omega$ scales with $r$ for fixed $u$ in this simple example where $ds^2$ is the physical metric and $d\tilde{s}^2$ is the metric for the unphysical spacetime). The condition on the stress tensor encompasses familiar asymptotic behavior for metrics for isolated matter sources.   The topology condition is needed to conclude that all asymptotically flat spacetimes are equipped with the same universal structure.  Namely, that of an equivalence class $[(q_{ab},n^a)]$  consisting of a degenerate metric $q_{ab}$ of signature $\{0,+,+\}$, and a null normal $n^a=g^{ab}n_b$ such that
\be
\mathcal{L}_n q_{ab}=0~~q_{ab}n^b=0
\ee
and where the equivalence class is formed by moding out by the rescaling
\be
(q_{ab},n^a)\mapsto (\omega^2 q_{ab},\omega^{-1} n^a)
\ee
when the conformal factor changes as $g_{ab}\mapsto \omega^2 g_{ab}$ such that
$\mathcal{L}_n \omega=0$.

We will not proceed further along this route, but pause to emphasize that such efforts enhance our understanding by distinguishing what structure is intrinsic to being asymptotically flat as compared to what data on top of this picks out a particular solution.
The data of a particular asymptotically flat spacetime is encoded in a connection on $\mathcal{I}$ that satisfies \be
D_a q_{bc}=0~~D_an^b=0,
\ee
and which can, for practical purposes, be induced from the bulk connection compatible with $g_{ab}$.  (The degeneracy of $q_{ab}$ is what makes the metric compatibility condition not unique.)  Pushing this analysis further, one can characterize the radiative modes in the fully non-linear theory in terms of data on $\mathcal{I}$.

\subsection{Bondi Expansion}\label{sec:bondi}
We will now introduce the coordinate based approach to defining asymptotically flat spacetimes, which will be the starting point of all of our analyses to come.  This program was spearheaded by Bondi, van der Burg, Metzner, and Sachs in the 60s (BMS)~\cite{doi:10.1098/rspa.1962.0161,doi:10.1098/rspa.1962.0206}.  The recent review~\cite{Compere} was useful to the summary here and should be of particular interest to those also wanting to learn more about covariant phase space techniques.

The first step is to define your coordinate system in some reasonable way, understand what physical data the gauge fixed metric components are encoding, and establish an asymptotic expansion in the region of interest (here either future or past null infinity).  The class of asymptotically flat spacetimes is then defined to be the set with a certain asymptotic form -- i.e. the space of solutions which obey the `appropriate' boundary conditions.  Identifying the appropriate boundary conditions is itself somewhat of an art.   While there are clearly certain solutions you do not want to exclude with too restrictive conditions, various researchers will sometimes loosen long-accepted ones, so their status is not so rigid.  In fact, most of the results of this thesis are based upon removing an otherwise reasonable restriction imposed by BMS that the sphere metric be non-singular.  While we have the insight of an intervening half-century to sharpen our hindsight, it is good to emphasize where things are in flux and that symmetries can be gained (see section~\ref{sec:Ward}).

With all this in mind, Bondi coordinates are defined as follows~\cite{Compere} (here we will focus on future null infinity).  Let a set of outgoing null radial geodesics be labeled by fixed $u$.  Then $n^\mu=g^{\mu\nu}\p_\nu u$ is the normal to this hypersurface, and the condition that this hypersurface is null  amounts to $g^{uu}=0$.  Labeling the transverse spacelike $S^2$ with $u$-independent coordinates  i.e. $n^\mu\p_\mu x^A=0$ implies $g^{uA}=0$. In lowered components, these translate to $g_{rr}=g_{rA}=0$.  The magnitude of the radial coordinate is then fixed to correspond to the luminosity distance, enforcing the inverse square law via $\p_r (\det(g_{AB})/r^2)=0$.  

One can use these above conditions to parameterize the most general Bondi gauge metric in the form~\cite{1106.0213BT}
\be
ds^2=e^{2\beta}\frac{V}{r}du^2-2e^{2\beta} dudr+g_{AB}(dx^A-U^A du)(dx^A-U^B du)
\ee
whereupon solving Einstein's equations allows one to simplify the functions parameterizing the metric components in terms of the free data you identify in this process.  For our purposes, we can start with the simplified expansion in terms of free data.

The flat metric in retarded coordinates $u=t-r$ also has the property that lines of fixed $(u,z,\bz)$ and varying $r$ are null.  Taking the large $r$ limit with the remaining coordinates fixed amounts to approaching future null infinity.  Thus we will be looking at metrics that approach 
\be
ds^2=-du^2-2dudr+2r^2\gamma_{z\bz}dzd\bz
\ee
as $r\rightarrow \infty$ and have components that can be written in an expansion in $\frac{1}{r}$.  The rate at which the subleading terms are suppressed as compared to the flat metric components in these coordinates involves a consistent interplay with the rate at which the matter stress tensor falls off near null infinity.  It turns out that for matter with the falloffs (see also~\cite{Campiglia2017, Pate2018, 1902.01840HMP} where one needs to start from this point if switching from Bondi gauge)
\begin{equation} \label{eq:Gfalloff1}
\begin{aligned}
T_{uu} \sim \mathcal{O}(r^{-2}), \ \ \ \ \ T_{ur} \sim \mathcal{O}(r^{-4}), \ \ \ \ \ T_{rr} \sim \mathcal{O}(r^{-4}), \\
T_{uA} \sim \mathcal{O}(r^{-2}), \ \ \ \ \ T_{rA} \sim \mathcal{O}(r^{-3}), \ \ \ \ \ T_{AB} \sim \mathcal{O}(r^{-1}),
\end{aligned} \end{equation}
one can write the metric near future null infinity as
\begin{align}
\label{eq:coord1}
ds^2 =&-du^2  -2du dr +2r^2 \gamma_{z\bz} dzd\bz \notag  \\ 
&+\frac{2m_B}{r}du^2 + rC_{zz}dz^2 +r C_{\bz\bz} d\bz^2 \notag \\
&+\left[(D^z C_{zz} -\frac{1}{4r} \p_z(C_{zz} C^{zz}) +\frac{4}{3r} N_z ) dudz +c.c\right]+...,
\end{align}
in terms of the free data 
\be\label{eq:free}
\{C_{zz}, m_B, N_z\}.
\ee
Note that here angular indices are raised with the unit sphere metric and $D_A$ denotes the covariant derivatives compatible with this metric.  Also $C_{\bz\bz}=C_{zz}^*$ and $N_{\bz}=N_z^*$ since the metric is real.  $C_{zz}(u,z,\bz)$ corresponds to the radiative data, and $N_{zz}=\p_u C_{zz}$ is referred to as the news tensor.  The news, but not $C_{zz}$, should decay to zero at early and late times in a system that is not continually radiating (up to a pure superrotation zero mode which we will discuss shortly).  $m_B$ is referred to as the Bondi mass and $N_z$ is the angular momentum aspect (where we have used the conventions of~\cite{1502.06120PSZ} for the definition of the angular momentum aspect in~(\ref{eq:coord1})).  Their $S^2$ averages would correspond to the total mass and angular momentum.  To compare this to the quantities from the ADM formulation, however,  one has to go to $u\rightarrow-\infty$ because their values change as a function of $u$ as radiation exits $\mathcal{I}^+$.  The $u$-evolution of $m_B$ and $N_z$ are given by the constraint equations which give the first order equations $G_{uu}=8\pi G T_{uu}$
\begin{align}\label{eq:cons1}
\pa_{u} m_{B} &={1 \over 4}  \left[D_z^2 N^{z z} + D^2_{\bar z} N^{\bar z \bar z} \right] - T_{uu}, \cr T_{u u} &\equiv {1 \over 4} N_{z z} N^{z z} + 4 \pi G \lim\limits_{r \to \infty} [r^2 T^{M}_{u u}] ,
\end{align}
 and $G_{u z} = 8 \pi G T_{u z}^{M}$
\begin{align}\label{eq:cons2}
\pa_{u} N_z &={1 \over 4}  \p_z \left[D_z^2 C^{z z} - D^2_{\bar z} C^{\bar z \bar z} \right] +\p_z m_B- T_{uz}, \cr
 T_{u z} &\equiv 8\pi G\lim\limits_{r\rightarrow\infty}[r^2 T^M_{uz}]-{1\over 4}D_z[C_{zz}N^{zz}]-{1\over 2}C_{zz}D_zN^{zz}.
\end{align}
Here we have grouped the quadratic terms with the matter stress tensor.  When analyzing inhomogeneous effects a lot can be gleaned from the much simpler linearized theory, and then corrected to include the higher order gravitational effects as another source term.

Now that we have expressed our metric expansion in powers of $r$ in terms of the free data, we can initiate an asymptotic symmetry analysis.   We look for vectors which keep the falloffs of~(\ref{eq:coord1}) invariant, and then ask how the free data transform under these diffeomorphisms.  Namely, we want
\begin{align}
\cL_{\xi} g_{ur} &=\co(r^{-2}),\;\;\;\;\cL_{\xi} g_{uz} =\co(1),\;\;\;\cL_{\xi} g_{zz} =\co(r) ,\;\;\;
\cL_{\xi} g_{uu} =\co(r^{-1}). 
\end{align}
We find that this is locally satisfied by the vector fields
\begin{align}\label{vf}
\xi  =&(1+\frac{u}{2r}) Y^{z}\p_z - \frac{u}{2r}  D^\bz D_z Y^{z} \p_{\bz} 
-\frac12 (u+r  )D_z  Y^{z} \p_r+{ u \over 2} D_z Y^{z} \p_u + c.c.\\ \notag
&+f\p_u  - \frac1r (D^zf\p_z + D^{\bz}f\p_{\bz}) +D^zD_zf\p_r +...,
\end{align}
parameterized by
\be
\{f(z,\bz), Y^z(z)\}
\ee
referred to as `supertranslations' and `superrotations' respectively (note the $f$ is real and $Y^\bz=(Y^z)^*$).  These $Y^A$ are conformal killing vectors (CKVs) of the celestial sphere metric.  These act non-trivially on the free data~\cite{1406.3312KLPS, He2015}
\be\begin{array}{rl}\label{es:srshift}
\delta_{\xi} C_{zz} =&  { u \over 2}( D_z Y^{z} +D_{\bz} Y^{\bz}) \p_u C_{zz}+\cL_{Y} C_{zz} -\frac12 ( D_z Y^{z} +D_{\bz} Y^{\bz})  C_{zz} - u D_{z}^3 Y^{z}\\
&+ f\p_u C_{zz}-2D_z^2 f
\end{array}\ee
and have finite but non-zero canonical charge~\cite{Crnkovic:1986ex,Henneaux:1992ig,Iyer:1994ys,Wald:1999wa} whose linear part is integrable~\cite{1106.0213BT}
\be\label{eq:qlin}
Q_{lin}=\frac{1}{8\pi G}\int \sqrt{\gamma} d^2 z \left((2f+uD_AY^A)m_B+Y^AN_A\right)
\ee
and evaluated at a cross section of future null infinity.  They are thus part of the asymptotic symmetry group~\cite{Strominger2017}
\be
ASG=\frac{\mathrm{Allowed~Gauge~Symmetries}}{\mathrm{Trivial~Gauge~Symmetries}}
\ee
where we quotient by those diffeomorphisms which fall off too fast to give a non-zero canonical charge, and the corresponding generators will be referred to as `large' gauge transformations.  We see that the set of asymptotic symmetries is much larger than Poincar\'e.  The extension of the translation subgroup of Poincar\'e to the infinite dimensional abelian supertranslations was noticed in the original efforts~\cite{doi:10.1098/rspa.1962.0161,doi:10.1098/rspa.1962.0206}.  The status of superrotations allowing arbitrary meromorphic $Y^z$ beyond the Lorentz generators
\be\label{alr}
\begin{array}{c}
 Y^z_{12} =iz,\;\;  Y^z_{13} = -\frac 12 (1+z^2) ,\;\; Y^z_{23} = -\frac i2 (1-z^2), \\
Y^z_{03} = z,\;\; Y^z_{01} =-\frac 12 (1-z^2),\;\; Y^z_{02} =-\frac i 2 (1+z^2),
\end{array}\ee
is central to this thesis.  
We will have more to say in section~\ref{sec:Ward} when we demonstrate the physical relevance of these symmetries to the perturbative gravitational $\mathcal{S}$-matrix.  There is still some more groundwork to cover first.  So far we have looked at the radial expansion of the metric.  We need to understand how the modes behave at early and late $u$ and how they match to data at past null infinity to make a statement about scattering.

\section{The Scattering Problem}\label{sec:scat}

In the last section, we introduced the asymptotic symmetry analysis near $\mathcal{I}^+$.  A similar story holds near $\mathcal{I}^-$.  However, one would only expect a diagonal subgroup of $BMS^+\times BMS^-$ to be a symmetry of scattering.  While we've identified the free data and form of the radial expansion, we need some matching between this data and the corresponding data at past null infinity to define the gravitational scattering problem.  

One of the key insights of Strominger was to propose an antipodal matching across spatial infinity~\cite{Strominger2014}.  Namely fields at $\mathcal{I}^+_-$ at $(z,\bz)$ get matched to fields at  $\mathcal{I}^-_+$ at $(z_A,\bz_{A})$ where the antipodal map on the Riemann sphere is
\be
z\mapsto -\frac{1}{\bz},
\ee
which indeed takes $\theta\mapsto \pi-\theta$ and $\phi\mapsto \phi+\pi$.  The original motivation cited the lefthand diagram of Figure~\ref{fig:penrose}.  In the conformal compactification, the generators of null infinity correspond to the lightcone of $i^0$, making an antipodal matching seem natural.  This is a singular point of the conformal compactification, so one should rightfully be cautious.  

One nice example from~\cite{Strominger2017} shows how a familiar smooth field configuration can exhibit this antipodal matching due to the order of limits implicit in going to $\mathcal{I}^+$ versus $\mathcal{I}^-$.  If we take the radial electric field of a relativistic charge with velocity $\beta$ and Lorentz boost factor $\gamma=\frac{1}{\sqrt{1-\beta^2}}$
\be
F_{rt}(t,r\hat{x})=\frac{e^2}{4\pi}\frac{Q\gamma(r-t\hat{x}\cdot\beta)}{|\gamma^2(t-r\hat{x}\cdot \beta)^2-t^2+r^2|^{3/2}}
\ee
then plugging in $t=u+r$ or $t=v-r$ and taking the $r\rightarrow\infty$ limit gives respectively
\be\begin{array}{ll}\label{eq:boostq}
F_{ru}(u,r\hat{x})&=\frac{e^2}{4\pi}\frac{Q\gamma(r-(u+r)\hat{x}\cdot\beta)}{|\gamma^2((u+r)-r\hat{x}\cdot \beta)^2-(u+r)^2+r^2|^{3/2}}\\
&=\frac{e^2}{4\pi r^2}\frac{Q}{\gamma^2(1-\hat{x}\cdot \beta)^2}+...\\
\end{array}\ee
while
\be\begin{array}{ll}
F_{rv}(v,r\hat{x})&=\frac{e^2}{4\pi}\frac{Q\gamma(r-(v-r)\hat{x}\cdot\beta)}{|\gamma^2((v-r)-r\hat{x}\cdot \beta)^2-(v-r)^2+r^2|^{3/2}}\\
&=\frac{e^2}{4\pi r^2}\frac{Q}{\gamma^2(1+\hat{x}\cdot \beta)^2}+...\\
\end{array}\ee
From the point of view of plane waves passing freely through spacetime, entering and exiting at antipodal points is also natural.  However the fact that this matching occurs between the data at $u\rightarrow-\infty$ and $v\rightarrow+\infty$ is non-obvious.  
For the moment we will take the pragmatic approach that this matching does work for establishing a Ward identity using the soft theorem.     

In what follows the antipodal matching will be implicit in our choice of $z$ coordinate near past null infinity.  This amounts to flipping the overall sign in the last two lines of~(\ref{flt}) near $\mathcal{I}^-$ to avoid the appearance of $z_A$'s in our matching conditions.  For our study of gravitational scattering the following matching conditions will be assumed~\cite{1406.3312KLPS,1502.06120PSZ}
\be\begin{array}{c}\label{eq:match} C_{zz}|_{\ci^+_-}=C_{zz}|_{\ci^-_+},~~~~~m_B|_{\ci^+_-}=m_B|_{\ci^-_+},\\
  \p_{[z}N_{\bz]} |_{\ci^-_+}= \p_{[z}N_{\bz]} |_{\ci^+_-}, \end{array}\ee
  at each $(z,\bz)$.
We will elaborate on further restrictions on the $u,v$ behavior as these fields approach these $\mathcal{I}^\pm_\pm$ limits as we need them in section~\ref{sec:Ward}, but emphasize here these also play an important part of defining what particular class of scattering problems are being considered.  

\subsection{Soft Theorems}\label{sec:soft}

We will need an additional field theoretic tool, that at first glance would appear unrelated to the story of asymptotic flatness and asymptotic symmetries we've begun to construct above: soft theorems.  Coincidentally the relevant theorems were established by Weinberg~\cite{Weinberg:1965nx} (see also~\cite{GellMann:1954kc,Low:1954kd,Low:1958sn,Burnett:1967km,Adler:1966gc,Gross:1968in}) around the same time as the analyses by BMS. 
The statement of these theorems amounts to the following observation:  given a gauge theory amplitude, the amplitude with one additional gauge boson exhibits a universal behavior as the momentum of the added gauge boson is taken soft.  Namely, it can be written in terms of a soft factor times the original amplitude without this extra boson.  

It is straightforward to prove the leading term diagrammatically, and a good textbook reference is chapter 13 of~\cite{WeinbergVol1} (see also chapter 6 of~\cite{Peskin:1995ev}).  Begin with an on-shell scattering amplitude $\mathcal{A}(p_i)$, and consider the related amplitude where we add an outgoing massless gauge boson to the final state, leaving the other scatterers unchanged. Taking the momentum of extra boson to be
\be\label{eq:q}
q^\mu=\omega (1,\hat{q})
\ee
we can ask what happens when we tune down the energy $\omega\rightarrow 0$.

 \begin{figure}[bth!]
\centering
\includegraphics[clip, trim=3cm 20.5cm 3cm 2.5cm, width=.8\textwidth]{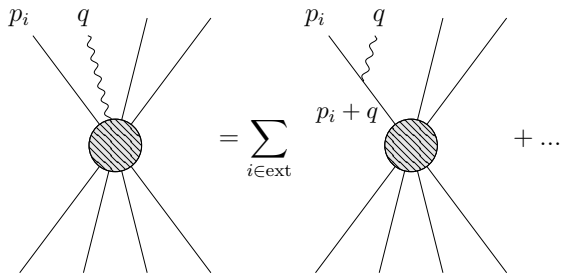}
 \caption{Diagrammatic expansion of an amplitude with an extra gauge boson as its energy is tuned towards zero.  The leading contributions come from the soft gauge boson attaching to the external lines.
 }\label{fig:softfactor}
\end{figure}

As in Figure~\ref{fig:softfactor}, there will be contributions to this amplitude that come from diagrams where the extra gauge boson attaches to one of the other external legs (which we will denote as `hard' particles in contrast to the `soft,' low energy added gauge boson).  Up to the vertex factor and an extra propagator, the remainder of the diagram will take the same form as $\mathcal{A}(p_i)$, up to the fact that one of the legs has momentum $p_i+q$ instead of $p_i$ (or $p_i-q$ when attached to an incoming leg) and so would no longer be on-shell.  However, in the limit that $\omega\rightarrow0$ this momentum approaches the on-shell $p_i$ again, and the extra propagator is proportional to
\be
\frac{-i}{(p_i+q)^2+m_i^2-i\e}=\frac{-i}{2p_i\cdot q-i\e}\propto \frac{1}{\omega}
\ee
(using $p_i^2=-m^2$ and $q^2=0$) which diverges in this limit.  (For a fermionic external leg the numerator of the Dirac propagator rewritten to have this denominator has a finite limit as $\omega\rightarrow0$.)  The other diagrams left out in the ellipses will not contribute to this `Weinberg pole.'  The form of the vertex factor will depend on the spin of both the added boson and  the external leg it is attaching to, however with some algebra one can show the latter dependence drops out when combined with the modified propagator numerators as a result of completeness relations for the appropriate spinors (or polarization tensors in the integer spin case)~\cite{WeinbergVol1}.  One can extend this derivation to $\mathcal{O}(\omega^0)$ and find the universal relation
\be\label{eq:soft}
\langle out| a_\pm(q) \mathcal{S}|in\rangle=(S^{(0)\pm}+S^{(1)\pm})\langle out| \mathcal{S}|in\rangle+\mathcal{O}(\omega)
\ee
where $\mathcal{A}(p_i)=\langle out|\mathcal{S}|in\rangle$ and
\be\label{eq:softem}
S^{(0)\pm}=e\sum_k  Q_k\frac{p_k\cdot\e^\pm}{p_k\cdot q}~~~S^{(1)\pm}=-ie \sum_k Q_k\frac{q_\mu\e^\pm_\nu J_k^{\mu\nu}}{p_k\cdot q}
\ee
for an additional soft photon, while

\be\label{eq:softgr}
S^{(0)\pm}=\frac{\kappa}{2}\frac{(p_k\cdot \e^\pm)^2}{p_k\cdot q}~~~S^{(1)\pm}=-i\frac{\kappa}{2}\sum_k\frac{p_{k\mu}\e^{\pm\mu\nu} q^\lambda J_{k\lambda\nu}}{p_k\cdot q}
\ee
for an additional soft graviton (where this form for the sub-leading soft theorems holds only at tree-level).  The $\pm$ subscripts indicate the helicity of the soft gauge boson and $\kappa=\sqrt{32\pi G}$.  Here we have assumed only one soft gauge boson has been inserted.  Also we are suppressing that the summation would involve a signed sum of the form `out'-`in,' if one takes $p^0_k>0$ for all legs --  i.e. in our conventions global energy-momentum and angular moment conservation imply $\sum p_k=0$ and $\sum J_{k\mu\nu}=0$ respectively (and for the $U(1)$ case $\sum Q_k=0$).  Note that the $J_{k\mu\nu}$ acts as a differential operator on $\mathcal{A}(p_i)$. In each case $S^{(0)}$ is $\mathcal{O}(\omega^{-1})$ and $S^{(1)}$ is $\mathcal{O}(\omega^0)$.  The soft graviton relation actually extends to $\mathcal{O}(\omega)$~\cite{1404.4091CS}. 
 
The leading soft theorems are due to Weinberg~\cite{Weinberg:1965nx}.  The subleading soft photon is due to the work of Low, Burnett, Kroll, Goldberger, and Gell-Mann~\cite{Low:1954kd, Low:1958sn, Burnett:1967km, GellMann:1954kc} (see also~\cite{Adler:1966gc, Casali:2014xpa, Bern:2014vva}).  The subleading soft graviton theorem was not discovered until nearly fifty years after the leading soft graviton theorem~\cite{1404.4091CS}, and the search for it was inspired by the emerging connections between existing soft theorems and asymptotic symmetries~\cite{Strominger2014, Strominger:2013lka}.
Followup papers by Strominger \& co. that have further explored these connections include~\cite{He2015,Strominger2016a, Kapec:2015vwa} for the leading soft graviton;~\cite{He2014, Kapec:2014zla, He:2015zea, Kapec2015, Strominger2016b, Nande2018, Pate:2017vwa, Ball:2018prg} for the leading soft photon and gluon; and~\cite{Lysov2014, Himwich:2019dug} for the subleading soft photon. The topic of this thesis will be the subleading soft graviton, which has the most recent origin story and is a prime example of benefits we gain from filling in gaps based on the framework of the IR triangle.
 
Now that we have stated the soft theorems for a single soft gauge boson, it is important to point out that this is only the tip of the iceberg.  The Weinberg poles appear in loop corrections to amplitudes, when a gauge boson attached to two external legs goes soft.  Here, the effect of the pole is more sinister because the loop momentum must be integrated over, giving a divergence in $d\le4$ dimensions.  When you sum over any number of such possible loops, this divergence exponentiates with the sign of the exponent such that amplitude is damped and vanishes when you remove your IR cutoff. 

One way to get a finite answer is to use the Weinberg pole divergence for soft emission to our advantage. When summing over an arbitrary number of additional soft photons or gravitons, one can show that the leading soft factors exponentiate.  Thus, if one argues that a `physical' observer could not measure boson emission at energies below some cutoff $E$ and missing total energy up to a scale of the same order of magnitude, then the IR cutoff dependence of the inclusive cross section cancels between virtual and real soft emissions, leaving only a dependence on the physical measurement scale $E$.  
Moreover in this soft regime, the exponential behavior of the cross section implies that soft gauge boson emissions follow a Poisson distribution, which has bearing in connecting to semiclassical interpretations of Bremsstrahlung (see sections 6.1 and 6.5 of~\cite{Peskin:1995ev}).  We will be interested in such connections in our study of memory effects in section~\ref{sec:mem}.

Thus one might fear that we have to abandon the notion of a well defined (IR finite) $\mathcal{S}$-matrix and settle for inclusive cross sections.  However, the work of Cheung, Kibble, and Faddeev and Kulish~\cite{Chung, Kibblei, Kibbleii, Kibbleiii, Kibbleiv, FK} suggested coherent state scattering as an alternative --  i.e. the fact that $\mathcal{S}$-matrix elements between the standard Fock states are zero is a symptom of considering the wrong scattering states rather than a diagnosis that no IR finite $\mathcal{S}$-matrix exists. 
This continues to be an active area of research, both from the point of view of asymptotic symmetries~\cite{Gabai:2016kuf,Mirbabayi:2016axw,Gomez:2017soz,Panchenko:2017lkw} and on the phenomenological side~\cite{Contopanagos:1990xe, Contopanagos:1991yb, Contopanagos:1992fm,Forde:2003jt,Frye:2018xjj}.  One interesting viewpoint promulgated by~\cite{Kapec:2017tkm} (see also~\cite{Strominger2017}) is that the vanishing of the Fock state $\mathcal{S}$-matrix elements is a consequence of non-conservation of the `large' gauge charges associated with the relevant asymptotic symmetry group.  The sum over soft states within inclusive cross sections gives a non-zero result because there is a state in that sum which satisfies the charge conservation condition.

\subsection{Mode Expansions}\label{sec:mode}
In order to make any statement connecting the asymptotic symmetry analysis of section~\ref{sec:asym} to the soft theorems studied here, we need to know  how to relate the radiative free data in~(\ref{eq:free}) to the creation/annihilation operator insertion in~(\ref{eq:soft}) and its incoming soft boson analog.  We will include the photon and graviton examples, with the latter being the primary focus in what follows.  Conventions used here for the photon case can also be found in~\cite{He2014}, and those for the graviton in~\cite{1406.3312KLPS}. Using bondi coordinates as in~(\ref{flt}) and a null boson momentum as in~(\ref{eq:q}) with energy $\omega_q$ and $\hat{q}$ parameterized by the projective coordinate $w$ we have
\begin{equation}\label{gravmom}
q^\mu = \frac{\omega_q}{1 + w {\bar w}} \left( 1 + w {\bar w} , w + {\bar w} ,  i \left( \bar{w} -  w\right), 1 - w {\bar w}  \right).
\end{equation}
The relevant polarization tensors are
\bea \label{pol1}
{ \e}^{+\mu}( {\vec q} ) &=& \frac{1}{\sqrt{2}} \left( {\bar w}, 1, - i, - {\bar w} \right),  \cr
{\e}^{-\mu}({\vec q} ) &=& \frac{1}{\sqrt{2}} \left( w , 1,   i, - w  \right),
\eea
for a $\pm 1$ helicity photon, and 
\bea
\e^{\pm\mu\nu} &=&\e^{\pm\mu}\e^{\pm\nu},
\eea
for a $\pm 2$ helicity graviton.  These obey $\e^{\pm\mu}q_\mu=\e^{\pm\mu}{}_\mu=0$.   We can then perform a mode expansion
\be
\mathcal{A}_\mu(x)=e\sum\limits_{\alpha=\pm} \int \frac{ d^3q}{(2\pi)^3} \frac{1}{2 \omega_q} \left[ \e^{\alpha*}_{\mu} ({ \vec q})a_\alpha ({\vec q}) e^{i q \cdot x} + \e^\alpha_{\mu}({\vec q}) a_\alpha ({\vec q})^\dagger   e^{- i q \cdot x} \right],
\ee
and
\begin{equation}
h_{\mu\nu}(x) = \sum\limits_{\alpha=\pm} \int \frac{ d^3q}{(2\pi)^3} \frac{1}{2 \omega_q} \left[ \e^{\alpha*}_{\mu\nu} ({ \vec q})a_\alpha ({\vec q}) e^{i q \cdot x} + \e^\alpha_{\mu\nu}({\vec q}) a_\alpha ({\vec q})^\dagger   e^{- i q \cdot x} \right],
\end{equation}
where the Fourier coefficients are promoted to operators in the quantum theory that obey
\be\label{rrd}
[a_\alpha ({\vec q}), a_\beta ({\vec{q'}})^\dagger ]= 2\omega_q\delta_{\alpha\beta}(2\pi)^3\delta^3 \left( {\vec q} - {\vec q}' \right).
\ee
We can now identify which modes correspond to the free data at null infinity from section~\ref{sec:asym}.  From the definition of the metric expansion~(\ref{eq:coord1}), the radiative data from~(\ref{eq:free}) becomes
\begin{equation}
C_{\bz\bz}(u,z,\bz) = \kappa \lim_{r\to\infty} \frac{1}{r} h_{\bz\bz}(u,r,z,\bz),
\end{equation}
while the familiar radiation mode from an accelerating charge that is $\mathcal{O}(r^{-1})$ in Cartesian coordinates becomes
\be\label{eq:ab}
A_\bz(u,z,\bz)=\lim\limits_{r\rightarrow\infty}\mathcal{A}_\bz(u,r,z,\bz).
\ee
We note that while Bondi gauge and harmonic gauge (which is more convenient for evaluating the soft theorems) differ in how they propagate the free data to subleading modes, this discrepancy will not affect our interpretation of the leading radiative modes. 
In each case, the Jacobian matrix elements $\p_\bz x^\mu$ are used to transform to Cartesian coordinates.  Note that for large $r$ and fixed $u$, the phase factors are rapidly oscillating
\be
e^{iq\cdot x}=e^{- i \omega_q u - i \omega_q r \left( 1 - \hat{q}\cdot\hat{x} \right) }
\ee
and the limit as $r\rightarrow\infty$ picks out the saddle point in the angular integral over $\hat{q}$ in the mode expansion, localizing to the modes for which $\hat{q}$ is parallel to $\hat{x}$.  As relevant to our study of memory effects in section~\ref{sec:mem}, this implies that the momentum of the radiation points in the same direction as the spatial vector from the source to the location of the observer seeing/feeling it at large $r$. 
In particular, in $(u,r,z,\bz)$ coordinates, the polarization tensors have angular components
\be \label{vv}\e_\bz^+ \left({\vec q} \right) = \p_\bz x^\mu\e^+_\mu \left( \vec{q} \right)= \frac{ \sqrt{2} r \left( 1 + z\bar{w} \right)}{ \left( 1 + z {\bar z} \right)^2 }   ,~~~~\e_\bz^- \left({\vec q} \right) = \p_\bz x^\mu\e^-_\mu \left( \vec{q} \right)  = \frac{ \sqrt{2} r { z} \left( { w} - { z} \right) }{ \left( 1 + z {\bar z} \right)^2 },\ee
so at the saddle point $\e_\bz^-$ and, from the complex conjugated expressions, $\e_z^+$ vanish.  We then get the simple relations
\be\label{eq:saddle1}
A_\bz=  - \frac{i e\hat{\e}_{\bz}^+}{8\pi^2}  \int\limits^\infty_0   d\omega_q \left ( a_- (\omega_q \hat{x}) e^{- i \omega_q u} -
 a_+ (\omega_q \hat{x})^\dagger e^{ i \omega_q u} \right),
\ee
and
\be
\label{eq:saddle2}
C_{\bz\bz} =-\frac{i\kappa}{8\pi^2}\hat{\e}_{\bz\bz}^{+}\int_0^\infty d\omega_q\left(a_-(\omega_q\hat{x})e^{-i\omega_qu}-a_+(\omega_q\hat{x})^\dagger e^{i\omega_qu}\right),
\ee
where we have introduced the radially rescaled and saddle-point evaluated polarization tensors
\be
\hat{\e}^{+ }_{\bz} =\frac{1}{r} \e_{\bz}^{+}  = \frac{\sqrt{2}}{1+z\bz},~~~ \hat{\e}^{+ }_{\bz\bz} =\frac{1}{r^2} \e_{\bz\bz}^{+}  = \frac{2}{(1+z\bz)^2}.
\ee
Similar expansions exist at past null infinity, and are needed to discuss incoming radiation.  From the final form of the mode expansions~(\ref{eq:saddle1}-\ref{eq:saddle2}), one can see that the soft limits of section~\ref{sec:soft} pick out the slowly-varying-in-$u$ modes.  We now have the tools we need to make asymptotic symmetry statements about the tree level $\mathcal{S}$-matrix using soft theorems.  In the next section we will use these tools to demonstrate the physical relevance of superrotations to the perturbative gravitational $\mathcal{S}$-matrix.

\section{Superrotations}\label{sec:superR}
The goal of this section is three-fold.  We will prove a Ward Identity for superrotations in~\ref{sec:Ward}, describe the physical observable they correspond to in~\ref{sec:mem}, and show that the same soft mode we use in both of those subsections also can be interpreted as a stress tensor for a putative 2D holographic dual in~\ref{sec:stress}.  These computations will be based on the results of~\cite{1406.3312KLPS},~\cite{1502.06120PSZ}, and~\cite{Kapec2017}, respectively. The following section will push this interpretation beyond the infrared and build the tools to map any 4D $\cal{S}$-matrix element to the proposed 2D dual.

\subsection{Ward Identity}\label{sec:Ward}
In this section we show that the tree level scattering matrix for quantum gravity possesses a Virasoro symmetry.  This is an infinite dimensional enhancement of the Lorentz subgroup of isometries of Minkowski space, and is an essential step to motivating a 2D holographic dual to scattering in 4D asymptotically flat spacetimes.  The following calculations will closely follow the original demonstration in~\cite{1406.3312KLPS}, which use as a starting point the previous results:
\begin{itemize}
\item  There exists a universal relation between scattering matrix elements with and without an additional low momentum graviton in the external states, through subleading order as the energy of this graviton is tuned to zero.  This is the subleading soft graviton theorem of Cachazo and Strominger~\cite{1404.4091CS}.
\item There is a proposed extension of the asymptotic symmetry algebra of asymptotically flat spacetimes beyond the original BMS algebra of Lorentz transformations and supertranslations~\cite{doi:10.1098/rspa.1962.0161,doi:10.1098/rspa.1962.0206}, which allows punctures on the celestial sphere, promoting the global $SL(2,\mathbb{C})$ Lorentz transformations to two copies of the Witt algebra.  These are known as superrotations and were motivated in~\cite{hep-th/0306074, 0303006dBS}.  In particular, we will need the canonical charge computations of~\cite{0909.2617BT, 1102.4632BT, 1106.0213BT}.
\end{itemize}
We will demonstrate that $\cs$-matrix elements satisfy a Ward identity for a diagonal subgroup of superrotations  acting at future and past null infinity.  Moreover, this Ward identity holds precisely because these $\cs$-matrix elements obey the subleading soft graviton theorem~\cite{1404.4091CS}.  This exercise thereby cements the physical role of superrotations.

This computation combines the tools we've set up in sections~\ref{sec:asym} and~\ref{sec:scat}.  Since we are proving a symmetry of the scattering problem, we will need to write the $\mathcal{I}^-$ versions of many of the above equations as well.   Performing a Bondi expansion of the metric near past and future null infinity we find~(\ref{eq:coord1})
\begin{align}
\label{eq:coord12}
ds^2 =&-du^2  -2du dr +2r^2 \gamma_{z\bz} dzd\bz \notag  \\ 
&+\frac{2m_B}{r}du^2 + rC_{zz}dz^2 +r C_{\bz\bz} d\bz^2 \\
&+\left[(D^z C_{zz} -\frac{1}{4r} \p_z(C_{zz} C^{zz}) +\frac{4}{3r} N_z ) dudz +c.c\right]+...\notag,
\end{align}
near $\mathcal{I}^+$ and 
\begin{align}
\label{eq:coord2}
ds^2 =&-dv^2+2dvdr +2r^2\gamma_{z\bz} dzd\bar{z} \notag\\
&+\frac{2m_B}{r}dv^2 - rC_{zz}dz^2 - r C_{\bz\bz} d\bz^2 \\
&+\left[(D^z C_{zz} +\frac{1}{4r} \p_z(C_{zz} C^{zz}) +\frac{4}{3r} N_z ) dvdz +c.c\right]+...\notag,
\end{align}
near $\mathcal{I}^-$.  The superrotation vector field was given near $\mathcal{I}^+$ in the first line of~(\ref{vf}) and, to distinguish the actions near $\mathcal{I}^\pm$ we add superscripts to the $Y^z$ here
\be \label{xip}
\xi^+  =(1+\frac{u}{2r}) Y^{+z}\p_z - \frac{u}{2r} D^{\bz} D_z Y^{+z} \p_{\bz} 
-\frac12 (u+r  )D_z  Y^{+z} \p_r+{ u \over 2} D_z Y^{+z} \p_u + c.c.+...
\ee
near $\mathcal{I}^+$ and
\be \label{xim}
\xi^-  =(1-\frac{v}{2r}) Y^{-z}\p_z +\frac{v}{2r}D^{\bz}D_z Y^{-z} \p_{\bz} 
-\frac12 (r-v  )D_z  Y^{-z} \p_r+\frac v2  D_z Y^{-z} \p_v + c.c.+... 
\ee
near $\mathcal{I}^-$.  The diagonal subgroup of $BMS^+\times BMS^-$ is the one generated by $Y^{+z}(z)=Y^{-z}(z)$ where we note that on $\mathcal{I}^-$ we are implicitly using the antipodal $z$ as compared to near $\mathcal{I}^+$ corresponding to the sign flip of $\hat{x}$ in~(\ref{flt}) discussed at the beginning of section~\ref{sec:scat}. 

As detailed in section~\ref{sec:bondi}, in addition to the radiative data~$C_{zz}$, we have the Bondi mass aspect $m_B$ and angular momentum aspect $N_z$ which appear in the canonical charge~\ref{eq:qlin}, and whose $u$-evolution is governed by the constraint equations~(\ref{eq:cons1}-\ref{eq:cons2}) 
\be\begin{array}{rl}\label{eq:con1}
\pa_{u} m_{B} &={1 \over 4}  \left[D_z^2 N^{z z} + D^2_{\bar z} N^{\bar z \bar z} \right] - T_{uu}, \cr T_{u u} &\equiv {1 \over 4} N_{z z} N^{z z} + 4 \pi G \lim\limits_{r \to \infty} [r^2 T^{M}_{u u}] ,\\~\\
\pa_{u} N_z &={1 \over 4}  \p_z \left[D_z^2 C^{z z} - D^2_{\bar z} C^{\bar z \bar z} \right] +\p_z m_B- T_{uz}, \cr
 T_{u z} &\equiv 8\pi G\lim\limits_{r\rightarrow\infty}[r^2 T^M_{uz}]-{1\over 4}D_z[C_{zz}N^{zz}]-{1\over 2}C_{zz}D_zN^{zz}, 
\end{array}\ee
near $\mathcal{I}^+$ and 
\be\begin{array}{rl}\label{eq:con2}
\pa_{v} m_{B} &=-{1 \over 4}  \left[D_z^2 N^{z z} + D^2_{\bar z} N^{\bar z \bar z} \right] + T_{vv}, \cr T_{vv} &\equiv {1 \over 4} N_{z z} N^{z z} + 4 \pi G \lim\limits_{r \to \infty} [r^2 T^{M}_{vv}] ,\\~\\
\pa_{v} N_z &=-{1 \over 4}  \p_z \left[D_z^2 C^{z z} - D^2_{\bar z} C^{\bar z \bar z} \right] +\p_z m_B+T_{vz}, \cr
 T_{v z} &\equiv 8\pi G\lim\limits_{r\rightarrow\infty}[r^2 T^M_{vz}]-{1\over 4}D_z[C_{zz}N^{zz}]-{1\over 2}C_{zz}D_zN^{zz},
\end{array}\ee
near $\mathcal{I}^-$.  In addition to the matching conditions in~(\ref{eq:match}), we need to specify the early and late $u$ and $v$ behavior of the radiative modes.  In these limits, we expect the matter stress tensor to vanish.
We will be following a generalization of the Christodouloud-Klainerman~\cite{ck} conditions that allows for superrotations.  Imposing the restrictions
\be\label{eq:ubc}
\p_\bz N_{zz}|_{\mathcal{I}^\pm_\pm}=0,~~~
D_\bz^2 C_{zz}-D_z^2 C_{zz}|_{\mathcal{I}^\pm_\pm}=0,
\ee
we consider
\bea 
\label{eq:asym}
C_{zz}(u,z,\bz)& \sim  -2 u   ({\p_zw})^{1/2} \p_z^2 ({{\p_zw})^{-1/2}}-2D_z^2f +\co(u^{-3/2}),\cr
C_{zz}(v,z,\bz)& \sim 2v  ({\p_zw})^{1/2} \p_z^2 ({\p_zw})^{-1/2} +2D_z^2f +\co(v^{-3/2}). \eea
The first term is the exponentiated version of the infinitesimal superrotation shift in~(\ref{es:srshift}), while the second term is a pure supertranslation which also appears there.

Let us refer to the linearized canonical charge~(\ref{eq:qlin}) defined at future null infinity as $Q^+$, and the corresponding expression at past null infinity as $Q^-$.  These are currently defined in terms of the bondi mass and angular momentum aspects at a given cross section of $\mathcal{I}^\pm$, which we take to be $\mathcal{I}^+_-$ and $\mathcal{I}^-_+$, respectively, so that we may apply our matching conditions of section~\ref{sec:scat}.  However, we can use the constraint equations~(\ref{eq:con1}-\ref{eq:con2}) and Stokes's theorem to rewrite $Q^+$ ($Q^-$) as an integral over future (past) null infinity plus a boundary term at $\mathcal{I}^+_+$ ($\mathcal{I}^-_-$).  In the following we assume only massless matter and that no black holes are formed, so that the boundary terms near $i^+$ and $i^-$ are trivial.  The generalization to the massive case has been considered in~\cite{Campiglia:2015kxa}.  In this case, we have the following superrotation charges
\be\begin{array}{ll}\label{eq:qlinpm}
Q^+(Y)&=-\frac{1}{8\pi G}\int \sqrt{\gamma} d^2 z du \p_u  \left(uD_AY^Am_B+Y^AN_A\right),\\
Q^-(Y)&=\frac{1}{8\pi G}\int \sqrt{\gamma} d^2 z dv \p_v  \left(uD_AY^Am_B+Y^AN_A\right).
\end{array}
\ee
Our goal will be to show that this charge implies a conservation law for perturbative $\cs$-matrix elements, namely
\be\label{dst}  \<out| Q^+(Y)\cs-\cs Q^-(Y)|in \>=0.\ee

One thing worth noting is that, as we mentioned above,~(\ref{eq:qlin}) is only the linearized charge and that there was a non-integrable term.  The obstruction in~\cite{1106.0213BT} is
\be
\Theta_s[\delta\chi,\chi]=\frac{1}{32\pi G}\int \sqrt{\gamma}d^2z (f+\frac{u}{2}D_CY^C)N_{AB}\delta C^{AB}
\ee
modulo addition of an exact term which is a total variation on phase space.  If this is evaluated at $\mathcal{I}^+_-$ then we can use~(\ref{eq:ubc}), to simplify $\Theta_s$.  This term would vanish if one considered only linearized variations around a background with no superrotation mode excited so that~$N_{zz}|_{\mathcal{I}^+_-}=0$ and not just $\p_\bz N_{zz}|_{\mathcal{I}^+_-}=0$.  This allows us to perform the supertranslation analysis~\cite{He2015} for which~(\ref{eq:qlin}) is the full charge when $Y^A=0$.  For superrotations, there is an additional integrable quadratic piece proportional to $D_AY^A$.  Here we take the approach of~\cite{1406.3312KLPS} and avoid this subtlety by accepting the fact that we can construct the charges~(\ref{eq:qlinpm}) starting from the linearized terms~(\ref{eq:qlin}) and show that~(\ref{dst}) holds.  In the end, the matching conditions of~(\ref{eq:match}) are enough to show that classically $Q^+=Q^-$, we now want to use the subleading soft theorem~(\ref{eq:softgr}) to show that this holds as an $\mathcal{S}$-matrix Ward identity.
  
 The righthand sides of the constraint equations~(\ref{eq:con1}-\ref{eq:con2}) take the form of a piece linear in $C_{AB}$  (recall $N_{AB}=\p_u C_{AB}$) plus a $T_{\mu\nu}$ term which is a combination of the matter stress tensor and terms quadratic in $C_{AB}$.  Inserting these expressions into the integral form of the charges~(\ref{eq:qlinpm}), one can see that this produces a natural splitting of $Q^\pm=Q_S^\pm+Q_H^\pm$ into a `soft' ($Q_S^\pm$) and a `hard' ($Q_H^\pm$) piece.  We refer to the term linear in the metric perturbation as the `soft' piece because, as we can see from the mode expansion of~(\ref{eq:saddle2}), the $u$ integral will pick up fourier modes with $\omega\rightarrow0$, in a manner we will make precise shortly.  Similarly, the `hard' piece picks up contributions from the matter fields and graviton modes that are not soft.  The terms quadratic in $C_{AB}$ that appear in $Q_H$ will act linearly on the graviton field, while the soft part contributes to an inhomogeneous shift (which is the origin of the Goldstone mode interpretation of these `large' gauge transformations~\cite{Strominger2014}).  
 
 We can then explicitly write down the soft part of the charge in terms of the mode expansion
 \be\label{eq:qs} Q^+_{S} = -\frac{1}{16\pi G} \int_{\mathcal{I}^+} du d^2z
  D_z^3  Y^{z} uN^z_{~\bz},~~~~Q^-_{S} =  \frac{1}{16\pi G} \int_{\mathcal{I}^-} dv  d^2z 
  D_z^3  Y^{z} vN^z_{~\bz}, \ee
 while the hard part is defined so as to act as $\xi^\mu\p_\mu$, appropriately transformed to momentum space.  For example,
 \be \label{oj}  Q^-_H|in\>=-i\sum_k \left( Y^{z}(z_k)\p_{z_k}-\frac {E_k}{2} D_zY^{z}(z_k) \p_{E_k}\right) |in\>, \ee
where the incoming momenta are parameterized by $(E_k,z_k,\bz_k)$
\be\label{eq:pmom}
p_k^\mu=\frac{E_k}{1+z_k\bar{z}_k}\left(1+z_k\bar{z}_k,\bz_k+z_k,i(\bar{z}_k-z_k),1-z_k\bar{z}_k\right).
\ee
 A similar expression exists for $Q_H^+$ acting on the $\langle out|$ state.
 Up to the caveat that the quadratic graviton part is added into our effective stress tensor, this would be the expected action of the matter stress tensor in the quantum theory sans gravity, as the generator of translations.  Here we have complexified the superrotation charge by considering only the $Y^z$ component.  One can add the Hermitian conjugated terms to get a proper superrotation preserving the reality of the metric, however this complexification will be useful when we construct a putative 2D stress tensor in section~\ref{sec:stress}.

 In order to regulate the $u$ integral appearing in $Q_S^+$, we define 
 \be
N^{\omega}_{\bz\bz}\equiv \int e^{i\omega u}  \p_u C_{\bz\bz} du,
\ee
where comparison with~(\ref{eq:saddle2}) shows that the sign of $\omega$ as $\omega\rightarrow0$ determines whether $a_-$ or $a_+^\dagger$ is selected.  Further defining the subleading soft graviton mode operator $N^{(1)}_{\bz\bz}$ as  
\begin{equation}\label{eq:n1}
\begin{array}{ll}
N^{(1)}_{\bz\bz}&\equiv\int duuN_{\bz\bz}\\&=-\lim\limits_{\omega\rightarrow0}\frac{i}{2}(\partial_\omega N^\omega_{\bz\bz}+\partial_{-\omega} N^{-\omega}_{\bz\bz})\\
&=\frac{i\kappa}{8\pi} \hat{\e}_{\bz\bz}^{+} \lim\limits_{\omega\rightarrow0}(1+\omega\partial_\omega)[ a_-(\omega\hat{x})-a_+(\omega\hat{x})^\dagger],\\
\end{array}
\end{equation}
we can check that the commutation relations~(\ref{rrd}) indeed imply that $Q_S^+$ as given in~(\ref{eq:qs}) generates the appropriate inhomogeneous shift~(\ref{es:srshift})
\begin{align} \label{eq:qbracket}
\scalemath{0.85}{[Q_S^+,C_{zz}]}&\scalemath{0.85}{={\frac{1}{32\pi^3}\int \sqrt{\gamma}d^2w D_w^3Y^w \lim\limits_{\omega\rightarrow 0}(1+\omega\p_\omega)\int_0^\infty d\omega_q ([a_+(\omega \hat{x})^\dagger, a_+(\omega_q\hat{x})]e^{-i\omega_q u}+[a_-(\omega\hat{x}),a_-(\omega_q\hat{x})^\dagger ]e^{i\omega_q u})}}\notag\\
&=\scalemath{0.85}{i D_z^3Y^z \lim\limits_{\omega\rightarrow 0}(1+\omega\p_\omega) \frac{\sin\omega u}{\omega}}\\
&=\scalemath{0.85}{i u D_z^3Y^z}\notag,
\end{align}
provided that $[Q_\xi,\cdot]\leftrightarrow-i\mathcal{L}_\xi$. 

Note that the $\lim\limits_{\omega\rightarrow 0}(1+\omega\p_\omega)$ projection operator that appears as a result of the $u\p_u$ in the $u$-integral of $C_{zz}$ projects out the leading Weinberg pole and any terms that vanish as $\omega\rightarrow0$ in~(\ref{eq:soft}), leaving us with precisely the subleading soft factor term at $\mathcal{O}(\omega^0)$.  To check the Ward identity~(\ref{dst}), we thus need to show the soft theorem is consistent with the hard action~(\ref{oj}), namely
\be\label{softmore} \<out| Q^+_S\cs-\cs Q^-_S|in \>=-\<out| Q^+_H\cs-\cs Q^-_H|in \>.\ee
Note that either an outgoing or an incoming soft graviton will give a soft factor with contributions from both the $in$-state and $out$-state  particles.  Meanwhile, the hard part of the charges act on each separately.  By 4D crossing the soft factor for an outgoing $-2$-helicity graviton is equal in magnitude, but opposite in sign, to an incoming $+2$-helicity graviton.  Thus the contributions from $Q_S^+$ and $Q_S^-$ in~(\ref{softmore}) add to twice the contribution of $Q_S^+$ coming from the annihilation operator term in $N^{(1)}_{\bz\bz}$ acting on $\langle out|$.  Explicitly our parameterization~(\ref{eq:pmom}) gives the soft factor~(\ref{eq:softgr})
\be\label{eq:s1w1}
S^{(1)-}=\scalemath{0.9}{\frac{\kappa}{2}\sum_k\left(\frac{E_k(z-z_k)(1+z\bar{z}_k)}{(\bz_k-\bz)(1+z_k\bar{z}_k)}\partial_{E_k}+
\frac{ (z-z_k)^2}{(\bz_k-\bz)}\partial_{z_k}+\sum_k \frac{(z-z_k)(1+z\bar{z}_k)}{(\bz-\bz_k)(1+z_k\bar{z}_k)}s_k\right)}
     \ee 
  where the $k^{th}$ particle has momentum $p_k$ and helicity $s_k$ (recall $J_{k\mu\nu}$ can be written as a differential operator acting on the on-shell particle momenta).   One then just needs to verify
\be
\gamma^{z\bar{z}}D_{z}^3(\hat{\e}_{\bz\bz}^{+}S^{(1)-})=-\pi \kappa\sum_k\bigl( D_{z}\delta^{(2)}(z-z_k)E_k\partial_{E_k}
+2\delta^{(2)}(z-z_k)\partial_{z_k}-s_k D_{z}\delta^{(2)}(z-z_k)\bigr)
\ee
and plug this into $Q_S^+$ with $N^{(1)}_{\bz\bz}\leftrightarrow \frac{i\kappa}{8\pi} S^{(1)-}$ to show 
\be\label{dki}
\begin{array}{ll}
\<out| Q^+_S\cs-\cs Q^-_S|in \>&= -i \sum\limits_k \scalemath{0.8}{\left( Y^z(z_k)\p_{z_k}-\frac{E_k}{2} D_zY^z(z_k) \p_{E_k}+\frac{s_k }{2} D_zY^z(z_k)\right)} \<out|\cs|in \>,
\end{array}
\ee
so that~(\ref{softmore}) is satisfied.  We have thus completed our first task, showing that superrotations are a relevant symmetry of the $\cs$-matrix in perturbative quantum gravity.  Note that while the subleading soft theorem implies the superrotation Ward identity, the converse is not true.  This has led to ongoing discussions of even further enlargement of the asymptotic symmetry group, e.g. from CKVs to Diff$(S^2)$~\cite{Campiglia2014}, which require relaxing the boundary conditions.
  Also, note the results here only hold at tree-level due to the validity of the soft theorem.  However, loop corrections to the subleading soft theorem have been investigated in~\cite{Bern:2014oka, He:2014bga, Bianchi:2014gla, Bern:2014vva, Broedel:2014bza} and~\cite{He:2017fsb} provides a one-loop exact correction to the soft graviton mode in $Q_S$ and the stress tensor we will define in~\ref{sec:stress}, so that it maintains the 2D Ward identity.  
 Before investigating what this soft graviton mode means in 2D, we will discuss its observable effects in 4D, namely as a new memory effect we call `spin memory'~\cite{1502.06120PSZ}.

\subsection{Spin Memory}\label{sec:mem}
The low energy fourier modes of the radiative data that appear as soft operators in $\cs$-matrix elements classically correspond to  physical observables that are long term effects.  The first step to appreciating what these observables are is to understand how to translate the gauge field fluctuations we've been studying into effects on test objects.  Then one can ask how to select the particular soft modes that correspond to a given soft factor.  Because we saw above that soft theorems are associated to asymptotic symmetries we see that these observables are also harbingers of the relevance of a certain asymptotic symmetry.  Put another way, we can see what exactly we expect to learn by measuring a particular memory effect.  

We would not expect their measurement to tell us something new about the underlying equations of motion, because they are derived from the Maxwell and Einstein equations we already know.  Rather, they do serve to tell us something about the appropriate boundary conditions.  For example, one with a very pragmatic bent might nix the study of asymptotic symmetries of asymptotically {\it flat} spacetimes from the start, given that current observations tell us that our universe has a non-zero cosmological constant.  So we should have really started with an analysis of asymptotically de Sitter spacetimes, which have a different conformal boundary structure.  The memory effects here serve as a bit of a reality check, bringing infinity from the `edge of spacetime' to the `radiation zone' relative to an astrophysical or particle collider event of interest -- where one is far away compared to the scale set by the impact parameter, but much smaller than the scales set by our tiny cosmological constant.

Our ultimate goal for this section is to understand the spin memory effect.  As a warm-up we will introduce the electromagnetic~\cite{Bieri:2013hqa, Tolish:2014bka} and leading gravitation displacement memory~\cite{1974SvA....18...17Z,Braginsky:1986ia, gravmem3, Ludvigsen:1989kg,Christodoulou:1991cr, Wiseman:1991ss, Thorne:1992sdb, Blanchet:1992br} analogs, in sections~\ref{sec:emem} and~\ref{sec:gdis}, following  \cite{Pasterski:2015zua} and \cite{Strominger2016a}.  We will then describe the spin memory effect in section~\ref{sec:spinmem}, and further discuss the interplay with asymptotic symmetries in section~\ref{sec:vactr}.

\subsubsection{A $U(1)$ Example}\label{sec:emem}
The leading electromagnetic example of the memory effect~\cite{Bieri:2013hqa} is one of the simplest to interpret and already pinpoints a contradistinction to arguments that use an effective detector cutoff, as in section~\ref{sec:soft}, to say that very low energy bosons won't be measured by real detectors~\cite{Pasterski:2015zua}.  The analog of the asymptotic symmetry analysis of section~\ref{sec:bondi}, is to perform a radial expansion of the gauge field on a fixed Minkowski space background and one finds that after appropriate gauge fixing the asymptotic symmetry group contains gauge transformations that modify $A_\bz$ in~(\ref{eq:ab}) by $\p_\bz \e(z,\bz)$~\cite{He2014}.  The point we care about here is that given an abelian gauge $\mathcal{A}_\mu$ field it is straightforward to evaluate its field strength $\mathcal{F}_{\mu\nu}$, and use the Lorentz force law to say how a test charge will react.  A non-zero leading Weinberg soft mode~(\ref{eq:softem})  implies that the time integral of the radiative part of the electromagnetic field is non-zero for generic scattering events.  This gives a net kick to freely floating test charges~\cite{Bieri:2013hqa}, a net displacement to test charges embedded in a viscous fluid~\cite{Pasterski:2015zua}, and the corresponding change in the gauge field may be measurable using superconductors~\cite{Susskind:2015hpa}.

In sum, Gauss's law applied to a Cauchy surface pushed up to null infinity relates the integral of the radiative field to the change between incoming and outgoing current configurations (as well as any massless charge fluxes through null infinity).  This correlates with the universality of the soft factor in that it only depends on the charge/kinematics of the asymptotic states.  Indeed, for massive charged matter, the leading soft theorem~(\ref{eq:softem}) obeys
\be\label{eq:ssf}\begin{array}{ll}
-\frac{e}{4\pi}\lim\limits_{\omega\rightarrow0}\omega [D^z\hat{\e}^{*+}_{z}S^{(0)+}+D^\bz  \hat{\e}^{*-}_{\bz}S^{(0)-}]&=-\frac{e^2}{4\pi}\sum\limits_k\frac{Q_k}{\gamma_k^2(1-\vec{\beta}_k\cdot\hat{x})^2},\\
\end{array}
\ee
for $p_k^\mu=m\gamma_k(1,\vec{\beta}_k)$.  The lefthand side is some operator acting on a low energy radiative mode ($\mathcal{O}(r^{-1})$ in Cartesian coordinates) while the righthand side is the superposition of the familiar Coulombic field (at $\mathcal{O}(r^{-2})$) of a boosted charge  as in~(\ref{eq:boostq})~\cite{Jackson:1998nia}.

As familiar from the  Li\'enard-Wiechert potential, a charge radiates when it accelerates, the time integral of its acceleration is just its velocity kick, so it's no surprise that a certain zero mode of the radiation only depends on the asymptotic trajectories of the charges.  This measurable effect on test objects depends on the magnitude of $\mathcal{F}_{\mu\nu}$ rather than its square, which would be the ordinary energy flux.  If one imagines controlling the acceleration as with charged beads moving on a rigid wire, we can tune the magnitude of the energy flux down arbitrarily by making the acceleration from the fixed initial trajectory to final trajectory take place over a longer time frame.  Thus even though we might not have a photodiode that will detect our soft quanta, we can still make observable statements about these low energy constraints on scattering.  

\subsubsection{Gravitational Displacement Memory}\label{sec:gdis} 
In the same manner that radiation from an accelerating charge will kick a test charge, a pulse of gravitational radiation displaces nearby inertial detectors.  This displacement memory effect has been studied since the 70s in \cite{1974SvA....18...17Z,Braginsky:1986ia, gravmem3, Ludvigsen:1989kg,Christodoulou:1991cr, Wiseman:1991ss, Thorne:1992sdb, Blanchet:1992br, Bieri:2011zb, Tolish:2014oda, Tolish:2014bka}.
The only barrier to making a similar statement about gravitational memory, is to know how to go from perturbations of the metric in Bondi gauge~(\ref{eq:coord1}) to statements about the motion of test bodies.  For this we can use the geodesic deviation equation.  If we consider two nearby inertial detectors sitting at large $r$ with tangent vector $t^\mu$, relative displacement vector $s^\mu$, and proper time $\tau$, we have
\be \p_\tau^2 s^\mu=R^\mu_{~\lambda  \rho\nu}t^\lambda t^\rho s^\nu.\ee
We can now evaluate the Riemann tensor for the metric expansion~(\ref{eq:coord1}), in particular 
\be {R_{zuzu}=-{1\over 2}r \p_u^2 C_{zz}.} \ee
Now because at large $r$ we can take $\tau\sim u$ and $t^\mu\p_\mu=\p_u$, we have
\be \p_u^2s^\bz={\gamma^{z\bz}\over 2r }\p_u^2C_{zz}s^z. \ee
We can integrate this twice in $u$ to find that along future null infinity the relative displacement changes by 
\be\label{eq:1overr}
\Delta^+ s^\bz=-\frac{\gamma^{z\bz}}{r}D_z^2\Delta^+ f s^z
\ee
at leading order in $\frac{1}{r}$.  We have used the $u$-falloff conditions~(\ref{eq:asym}) and the fact that the pure superrotation mode is projected out by the two $u$ derivatives.  (Note~\cite{Strominger2016a} and earlier analyses would have excluded the pure superrotation mode in~(\ref{eq:asym}),  and there are only non-trivial vacuum transitions in this mode for backgrounds with snapping cosmic strings~\cite{Strominger:2016wns}, so it would be pure gauge here.  We will have more to say about vacuum transitions in section~\ref{sec:vactr}.)  The point of~\cite{Strominger2016a} is to tie this memory effect into the asymptotic symmetry $\Leftrightarrow$ soft theorem results~\cite{Strominger2014,He2015}, by showing that the low energy mode picked up by the leading soft graviton theorem~(\ref{eq:softgr}) is precisely this displacement memory shift -- i.e. the Weinberg pole is the fourier transform of the relative displacement profile which looks like step-function at long time scales. 

There are various contributions to the memory effect~\cite{Compere}, in particular Christodoulou showed that gravitational radiation can itself source this displacement memory~\cite{Christodoulou:1991cr}.  A simple analog of~(\ref{eq:ssf}) holds for the contribution of a set of boosted massive particles, namely~\cite{Pasterski:2015zua}
\be\begin{array}{ll}
-\frac{\kappa}{16\pi}\lim\limits_{\omega\rightarrow0}\omega [D^z D^z \hat{\e}^{*+}_{zz}S^{(0)+}+D^\bz D^\bz \hat{\e}^{*-}_{\bz\bz}S^{(0)-}]&=\sum\limits_k \frac{Gm_k}{\gamma^3_k(1-\vec{\beta}_k\cdot\hat{x})^3}\\
\end{array}
\ee
where the lefthand side is the leading soft graviton factor~(\ref{eq:softgr}) and the righthand side $\sum m_B(m_k,\vec{\beta}_k)$ a sum of the Bondi mass contributions from a bunch of boosted objects, which can be compared to the appendix of~\cite{doi:10.1098/rspa.1962.0161}.

\subsubsection{A New Gravitational Memory}\label{sec:spinmem}
The result of~\cite{Strominger2016a} completed one iteration of the IR triangle.  With the above $U(1)$ case also understood in this context~\cite{Pasterski:2015zua}, it was clear that more iterations abounded.  We will now show that one can construct a memory effect corresponding to the subleading soft theorem. 

For the moment we will suppress the pure superrotation gauge mode in our analysis, and consider pure supertranslated early and late $u$ behavior~(\ref{eq:asym})
\be\label{eq:2vac}
C_{zz}|_{\mathcal{I}^+_\pm}=-2D_z^2 f^\pm.
\ee
As can be seen from the subleading soft mode $N_{\bz\bz}^{(1)}$ in~(\ref{eq:n1}), the $u\p_u C_{zz}$ appearing in the integrand projects out the pure supertranslated asymptotes of $C_{zz}$.  Another way to project out the $\mathcal{I}^+_\pm$ limits of $C_{zz}$, and thereby the contribution associated to the leading soft factor, is to use the fact that $f\in \mathbb{R}$, or the boundary condition this came from, namely~(\ref{eq:ubc})
\be\label{eq:round}
D_\bz^2 C_{zz}-D_z^2 C_{zz}|_{\mathcal{I}^+_\pm}=0.
\ee
In particular, this combination will have a finite $u$-integral, and we can use this to form our desired memory effect.

Ideally a `memory effect' is an observable that is not just low energy, but rather can be measured as a change between an initial state and final state, that can be measured by comparing an apparatus at early and late times.  In practice even though the leading memory effect has this form, that type of wait-and-repeat measurement would be more viable for future space-based detectors like eLISA.  Another proposal is to uses pulsar timing arrays~\cite{Pshirkov:2009ak} where the signature of the memory is a linearly growing pulsar timing residual coming from the time integral of a constant displacement in $C_{zz}$. (When a gravitational wave propagates between us and the pulsar its frequency shifts by a fraction proportional to $C_{zz}$ and the accumulated shift up to a given observation time is called the pulsar timing residual. While we want to project out such a linearly growing piece to get a finite spin memory, it is a feature that would help to observationally identify the displacement memory using this method.)  Yet another hope is to see the leading displacement memory in signals from Advanced LIGO~\cite{Lasky:2016knh}.  For this proposal one needs to process the waveform and take into account that the detector will be insensitive to very low frequency oscillations $\lesssim10$ Hz.  Given this variety for how to actually measure the displacement memory effect, we will content ourselves with still using the term `memory effect' to describe an experiment where we can compare a measurement at the beginning and end of a long integration time even if the experiment has to be constantly running during that interval.
 
We define the spin memory observable to be an accumulated time delay measured by a Sagnac-like detector where two counterrotating beams acquire this delay due to angular momentum flux of either gravitational waves or other spinning matter that passes through the ring.  Ordinary Sagnac interferometers are used for laser-based gyroscopes to measure rotation, and LIGO is a zero-area Sagnac interferometer~\cite{Sun:1996bj}.   Here we want the system to be sensitive to rotations.  Moreover we consider a configuration with the beam paths at fixed Bondi coordinates -- `BMS detectors' as used in~\cite{Strominger2016a}.  That an inertial ring would tend to rotate is consistent with the usual implementation of a  Sagnac detector (as there, one could keep track of the interference pattern between the two beams rather than when packets cross a moving reference point on the rotating ring).   The feasibility of measuring spin memory with the Einstein telescope has been considered in~\cite{Nichols:2017rqr}.

Consider such a detector whose path defines a circle $\mathcal{C}$ of radius $L$ centered at a point $z_0$ in Bondi coordinates near $\mathcal{I}^+$, i.e. for angular coordinate $\phi$ around this path

\be
Z(\phi)=z_0\left[1+{{Le^{i\phi}}\over {2r}}{{1+z_0\bar{z}_0}\over{\sqrt{z_0 \bar{z}_0}}}\right]+\CO\left({L^2\over r^2}\right).
\ee
A light ray that begins at $\phi=0$ at $u=0$ will follow a trajectory $\phi(u)$ such that 
\be\begin{array}{ll}
0=ds^2&=1-2r^2\gamma_{z \bar{z}}\p_u Z\p_u\bar{Z}-2{m_{B}\over r}-rC_{zz}(\p_u Z)^2-rC_{\bar{z} \bar{z}}(\p_u \bar{Z})^2\\&~~~~~~~~-[D^zC_{zz}\p_u Z+D^{\bar{z}} C_{\bar{z}\bar{z}}\p_u \bar{Z}]+...
\end{array}
\ee
As long as $C_{zz}$ does not change much during the time it takes to make a single orbit, the difference in time it takes for the two counter-propagating beams to each complete a single orbit will be given by the part that is odd under $\p_u Z\rightarrow -\p_u Z$
\be
\Delta P=\oint_\CC \bigl( D^{z} C_{z z}dz + D^{\zb} C_{\zb \zb}d\zb\bigr).
\ee
This vanishes for a pure supertranslation, as can be seen by invoking Stokes's theorem to rewrite this as a surface integral of the quantity~(\ref{eq:round}).   Now just as one would use the limit of Riemann sums to form the integral of an appropriately behaved function, measuring the accumulated time delay along future null infinity by adding up $\Delta P$ for each orbit gives the integral
\be\label{eq:deltau}
\Delta^+ u={1 \over 2\pi L }\int  du \oint_\CC \bigl( D^{z} C_{z z}dz + D^{\zb} C_{\zb \zb}d\zb\bigr),
\ee
which we will show is precisely the subleading soft graviton mode.

Using that the linearized expectation value of the metric perturbation obeys the semiclassical relation
\be\begin{array}{ll}\label{eq:spmem}
(\lim\limits_{\omega \to 0+}+\lim\limits_{\omega \to 0-})h_{\alpha\beta }(\omega,q)&= \epsilon_{\alpha\beta}(\lim\limits_{\omega \to 0+}+\lim\limits_{\omega \to 0-}){  \CA_{n+1}\bigl(p_1,...p_n;  (\omega q ,\epsilon_{\mu\nu})\bigr)\over\CA_{n}\bigl(p_1,...p_n\bigr)} \\
&=  ~{i\kappa} \epsilon_{\alpha\beta}\epsilon^{\mu \nu} \sum\limits_{k=1}^n{p_{k\mu} J_{k\nu\lambda} q^\lambda \over q\cdot p_k},
\end{array}\ee
where the symmetrized limits for $\omega\rightarrow 0^\pm$ is designed to cancel the leading Weinberg pole, the stationary phase approximation from section~\ref{sec:mode} then gives
\be
\int du C_{zz}(u,\hat q)-\int dv C_{zz}(v,\hat q) = -(
\lim_{\omega \to 0+}+\lim_{\omega \to 0-}){i \kappa\over 8\pi }\p_zX^\mu\p_z X^\nu h_{\mu \nu} (\omega, \hat q ) ,
\ee
so we find
\be\im\big[ \int du D^z D^z C_{\bz\bz}- \int dvD^z D^z C_{\bz\bz}\big]={\kappa\over 8\pi}[D^z D^z\hat{\e}^{*+}_{zz}S^{(1)+}-D^\bz D^\bz \hat{\e}^{*-}_{\bz\bz}S^{(1)-}].\ee
Up to an application of Stokes's theorem this is just the combination of past and future null infinity contributions of~(\ref{eq:deltau}).

We will now write an expression for the time delay~(\ref{eq:deltau}) in terms of a change in the angular momentum aspect and angular energy momentum flux through null infinity.  Multiplying the constraint equation~(\ref{eq:cons2}) by the Green's function
\be
{\cal G}(z;w) =\log\sin^2 { \Theta \over 2} , ~~~ \sin^2 {\Theta(z,w) \over 2} \equiv{|z-w|^2\over (1 + w \bar w)(1 + z \bar z) }
\ee
which obeys
\be{\p_z\p_\bz{\cal G}(z;w)=2\pi\delta^2(z-w)-{1\over 2}\gamma_{z\bz},}\ee
we find
\be
\Delta^+ u=-{ 1\over \pi^2 L}\im \int_{D_{\CC}} d^2w\gamma_{w\bar{w}}\int d^2z\p_\bz{\cal G}(z;w)\left[\Delta^+ N_z+\int du T_{uz}\right].
\ee
By considering only massless matter, as in section~\ref{sec:Ward} with trivial data at $i^\pm$, we have no contributions from $N_z|_{\mathcal{I}^+_+}$ and $N_z|_{\mathcal{I}^-_-}$, and can match the remaining across $i^0$ using~(\ref{eq:match}), to find a total time delay sourced by angular momentum flux through null infinity
\be\Delta \tau \equiv \Delta^+u-\Delta^- v=-{ 1\over \pi^2 L}\im \int_{D_{\CC}} d^2w\gamma_{w\bar{w}}\int d^2z\p_\bz{\cal G}(z;w)\left[\int du T_{uz}
-\int dv T_{vz}\right].
\ee

To get a sense of how small this effect is let us consider quadrupole radiation~\cite{Wiseman:1991ss} with no incoming news or stress tensor flux along $\mathcal{I}^-$ and see what its contribution to the above time delay will be.  The news tensor is given in terms of the $Y^z_{k}\propto\e_{ijk}Y^{z}_{ij}$ of~(\ref{alr})  
\be
N_{zz}= NY^i_zY^j_z
\ee
picking $i=j, Y^z_i=z$ and a gaussian profile in $u$
\be
N={\alpha \over (2\pi)^{1/4}}\p_u e^{-u^2+i\omega u},
\ee
we then have 
\be\int du T_{uz}={3i\over 2}\alpha^2\omega \p_z{z^2\bar{z}^2\over (1+z\bar{z})^4},\ee
so that the accumulated time delay around a contour $\mathcal{C}$ is
\be \Delta^+ u={3\alpha^2 \omega \over \pi L} \int_{D_{\CC}} d^2w\gamma_{w\bar{w}} \left[{w^2\bar{w}^2\over (1+w\bar{w})^4}-{1\over30}\right].\ee
If the area $D_{\CC}$ on the celestial sphere bounded by the curve $\mathcal{C}$ is of order $L^2/r^2$, we would have $\Delta^+ u \sim {\alpha^2 L \over r^2}$, which is to be compared with the $\frac{1}{r}$ scaling of~(\ref{eq:1overr}).

\subsubsection{Vacuum Transitions}\label{sec:vactr}
There is a stronger interplay between memory effects and asymptotic symmetries than we have emphasized thus far.  The examples have focused on how the subleading soft theorems correspond to memory observables.  Their connection to the asymptotic symmetries then hinges on the existing asymptotic symmetry $\Leftrightarrow$ soft theorem leg of the IR triangle.  However, there is more direct connection in terms of vacuum transitions and symplectic pairing which we now explore.

For the leading $U(1)$ and gravitational cases, we have early and late $u$ data related to one another by a large gauge transformation.  Avoiding superrotations for the moment, we have~(\ref{eq:2vac}) as the asymptotic behavior of $C_{zz}$ near ${\mathcal{I}^+_\pm}$.
Meanwhile performing a supertranslation generates the inhomogeneous shift
\be
\delta_{f}^{shift} C_{zz} =-2D_z^2 f
\ee
so that under a supertranslation the asymptotic `pure supertranslation' modes of the metric, parameterized by $f^\pm$ from~(\ref{eq:2vac}) shift as
\be
f^+\mapsto f^++f,~~~~f^-\mapsto f^-+f.
\ee
The memory mode~(\ref{eq:1overr}) depends only on the difference $f^+-f^-$ which is invariant to this large gauge transformation, as required for a physical observable.  However, the fact that the memory effect is non-zero is exactly why we can't try to trim down our phase space to obey more restrictive boundary conditions that would eliminate $f^\pm$ entirely.  We find that the leading memory effect tells us that typical scattering processes will induce dynamical vacuum transitions between supertranslation-labeled vacua. 

However, if it were just that the memory is supposed to be a vacuum-to-vacuum transition for the corresponding asymptotic symmetry group, something would seem out of place in the superrotation example.  Transitions between different superrotated vacua (first term in~(\ref{eq:asym})) correspond to spacetimes with snapping cosmic strings~\cite{Strominger:2016wns}.  But we know that the spin memory observable comes from a non-zero subleading soft theorem~(\ref{eq:spmem}), which involves much more mundane scattering processes.

In~\cite{1902.01840HMP} it was shown that the subleading soft graviton mode can be recast as a boundary difference of the metric component, in harmonic gauge.  We perform the linearized analysis here.  While the above statements were made in Bondi gauge, it is straightforward enough to compare expressions in a different gauge choice, by first rewriting the Bondi mass and angular momentum aspect in terms of components of the Weyl tensor.  Away from matter sources we have
\be\begin{array}{c}\label{eq:weylq}
\lim\limits_{r\rightarrow\infty} r\gamma^{z\bz}C_{u\bz zr}=-m_B-\frac{1}{4}C_{zz}N^{zz}+\frac{1}{4}(D^z D^z C_{zz}-D^\bz D^\bz C_{\bz\bz})\\
  \lim\limits_{r\rightarrow\infty}r^3C_{zrru}=N_z\end{array}\ee
which correspond to the Weyl scalars $\Psi_2^0$ and $\Psi_1^0$ of the Newman-Penrose formalism~\cite{Newman1962}, up to a rescaling due to tetrad normalization.  We can use this and the boundary conditions~(\ref{eq:round}) at $\mathcal{I}^+_\pm$ to write the charge for the linearized theory~(\ref{eq:qlin}) in terms of these Weyl tensor components.  

Considering linearized perturbations around Minkowski space $g_{\mu \nu} = \eta_{\mu \nu} + h_{\mu \nu} - \frac{1}{2}\eta_{\mu\nu}\eta^{\alpha\beta}h_{\alpha\beta}$ where $\eta_{\mu \nu}$ is the flat metric and $h_{\mu\nu}$ is the trace-reversed perturbation, the harmonic gauge condition reads 
\begin{equation}\label{eq:harga} \vspace*{-.1cm}
\nabla^{\mu}h_{\mu\nu}=0, \vspace*{-.1cm}
\end{equation} 
and the linearized Einstein equations reduce to
\begin{equation} \vspace*{-.1cm}
\square h_{\mu\nu} = - 16 \pi G T_{\mu\nu} .
\end{equation}
The same game of determining what falloffs are needed to consistently solve matter with falloffs~(\ref{eq:Gfalloff1}) are performed as in section~\ref{sec:bondi} in Bondi gauge, where we note that logarithmic-in-$r$ modes are a necessary feature in harmonic gauge.  After residual gauge fixing detailed in~\cite{1902.01840HMP}, one still has supertranslations and superrotations as part of the asymptotic symmetry group.  Moreover, we learn something about the subleading soft graviton mode by looking at the sphere metric one order subleading in $r$.  The analog of~(\ref{eq:asym}) allowing for an infinitesimal superrotation but no snapping cosmic strings is
\begin{equation}\label{eq:latetime}
  h_{\bz\bz,\pm}^{(-1)} =-uD_\bz^3{Y}^\bz(\bz) -2 D_\bz^2 {f}^{\pm}(z,\bar{z}), \ \ \ h_{\bz\bz,\pm}^{(0)} = u [D^2 - 2] D_\bz^2 {f}^{\pm}(z,\bar{z}) + 2D_\bz {V}_\bz^{\pm}(z,\bar{z}),
\end{equation}
where the early and late time behaviors of the metric are parameterized by an infinitesimal superrotation, supertranslation, and a {\it subleading} diffeomorphsim $\xi^\mu\p_\mu=r^{-2}V^A\p_A+...$  (Here for a field $\mathcal{O}$ we use $\mathcal{O}^{(n)}$ to denote the mode that scales like $r^{-n}$ in an expansion at large $r$.)  Evaluating the Weyl tensor components in this gauge, using~(\ref{eq:weylq}) to plug into the charge~(\ref{eq:qlin}), and separating this into soft and hard parts as in section~\ref{sec:Ward}, one finds that the soft charge simplifies to 
\be
Q_S^+(Y^z,Y^\bz=0)=\int \sqrt{\gamma}d^2z Y^z D_zD_z {V}^z\Big{|}^{\mathcal{I}^+_+}_{\mathcal{I}^+_-},
\ee
and since we saw that this contained the subleading soft graviton mode, we should also be able to find a nice expression for the spin memory effect in terms of $V^A$.  Indeed~(\ref{eq:deltau}) becomes
\be\label{eq:spinresult}
\Delta^+ u=\frac{1}{2\pi L}  \oint_{\mathcal{C}} {V}_A dx^A \Big{|}^{\mathcal{I}^+_+}_{\mathcal{I}^+_-} .
\ee
We see that spin memory is a proper `memory effect' to those who adhere to the stricter definition of a  wait-and-repeat rather than continuously accumulated effect~\cite{Compere}.

So for the superrotation case, we see that we can find a vacuum transition interpretation of the subleading soft graviton mode, however it is sub-radiative by one power of $r$.  This is reminiscent of the leading soft graviton mode in higher dimensions~\cite{Kapec:2015vwa}, where the analog of a memory effect appears at deeper orders in $\frac{1}{r}$.  The memory effect appears in the Coulombic mode at order $r^{3-d}$ which happens to coincide with the radiative mode, at $r^{1-\frac{d}{2}}$, when $d=4$~\cite{Pate2018}. (For more on the subtleties of the memory observable and the other side of the debate about the relevant asymptotic symmetry group in higher dimensions see~\cite{Hollands:2003ie, Tanabe:2011es, Hollands:2016oma, Garfinkle:2017fre}).

In this case, we see that while the spin memory has a nice expression in terms of a difference between $V_A^+$ and $V_A^-$, the subleading diffeomorphism that would generate the simultaneous shift
\be
V_A^+\mapsto V_A^++V_A,~~~~V_A^-\mapsto V_A^-+V_A
\ee
has vanishing canonical charge.  However, the bracket~(\ref{eq:qbracket}) should be a hint at a common interpretation of all three examples we have considered in this section.  Namely the relevant soft modes are symplectically paired to the pure `large-gauge' Goldstone modes.  This is in the same way one expects the constraint equation to generate a gauge transformation.
  This pairing has been worked out in~\cite{He2015} for the supertranslation case, a similar analysis of zero mode brackets for the $U(1)$ case can be found in~\cite{He2014}, and~\cite{Donnay:2018neh} has more to say about the superrotation case in the language of the conformal basis we will introduce in section~\ref{sec:cb}.

\subsection{A 2D Stress Tensor}\label{sec:stress}
Now that we have described memory effects and in particular the 4D interpretation of the subleading soft graviton mode as the spin memory effect, we conclude our discussion of superrotations with a summary of the results of~\cite{Kapec2017} which provides a very curious 2D interpretation of this same mode.  Namely they define an operator $T_{zz}$ in terms of the 4D soft graviton mode such that its insertion in 4D $\mathcal{S}$-matrix elements takes the same form as the OPE of a 2D stress tensor and a conformal primary when the 4D $\cs$-matrix element is viewed as a 2D celestial sphere correlator. The bulk of section~\ref{sec:cb} will be to make this map precise.  In the meantime, the groundwork we have covered in the previous sections puts us in a position where we are nearly ready to quote the results of~\cite{Kapec2017}, since we have already made explicit our convention for the metric~(\ref{eq:coord1}), mode expansions~(\ref{eq:saddle2}),~(\ref{eq:n1}), and soft theorem~(\ref{eq:softgr}).

The key insight is that the subleading soft graviton theorem, written out explicitly in~(\ref{eq:s1w1}) and~\cite{1406.3312KLPS} can be recast into the more 2D-covariant form
\be
S^{(1)-}=\frac{\kappa}{2}\sum_k\frac{(z-z_k)^2}{\bz-\bz_k}\left[\frac{2h_k}{z-z_k}-\Gamma^{z_k}_{z_k z_k}h_k-\p_{z_k}+|s_k|\Omega_{z_k}\right]
\ee
and similarly for $S^{(1)+}$ in terms of the operators
\be
h_k\equiv\frac{1}{2}(s_k-\omega\p_{\omega_k}),~~~\bar{h}_k\equiv\frac{1}{2}(-s_k-\omega\p_{\omega_k})
\ee
where $\Gamma^{z}_{zz}$ is the Christoffel connection for the unit round sphere metric~(\ref{eq:spheremetric}) and $\Omega_z=\frac{1}{2}\Gamma^{z}_{zz}$ is the spin connection.   In terms of $N^{(1)}_{\bz\bz}$ of~(\ref{eq:n1}) 
\be
T_{zz}\equiv\frac{i}{8\pi G}\int d^2 w\frac{1}{z-w}D_w^2 D^\bw N_{\bw\bw}^{(1)}.
\ee
Then provided we can construct a map from 4D $\cs$-matrix elements to 2D correlators such that
\be\label{eq:map42}
\langle out| \mathcal{S}|in\rangle\mapsto\langle \mathcal{O}_1...\mathcal{O}_n\rangle
\ee
for an $n$-particle scattering process, the soft theorem~(\ref{eq:soft}) turns an insertion of $T_{zz}$ into
\be\label{eq:TOPE}
\langle T_{zz}\mathcal{O}_1...\mathcal{O}_n\rangle=\sum_k^n\left[\frac{h_k}{(z-z_k)^2}+\frac{1}{z-z_k}(\p_{z_k}+h_k\Gamma^{z_k}_{z_k z_k}-|s_k|\Omega_{z_k})\right]\langle \mathcal{O}_1...\mathcal{O}_n\rangle
\ee
which is the expected form of a CFT stress tensor correlator on a curved background~\cite{Eguchi:1986sb}.  We note that this $T_{zz}$ is none other than the soft charge~(\ref{eq:qs})  for a particular superrotation
\be
T_{zz}=2i Q_{S}^+(Y^w=\frac{1}{z-w},Y^\bw=0)
\ee
and the charge for an arbitrary complexified superrotation can be constructed from a contour integral of this, using
\be
g(w)=\oint_{\mathcal{C}}\frac{dz}{2\pi i}\frac{g(z)}{z-w}
\ee
to build up an arbitrary superrotation from the one parameterized by a simple pole, i.e.
\be
Q_{S}^+(Y^w(w),Y^\bw=0)=-\frac{1}{4\pi}\oint_\mathcal{C} dz Y^z T_{zz}.
\ee  So the soft part of the superrotation charge corresponds to the current $j_z=Y^zT_{zz}$ in the putative CFT$_2$. We find from this computation that the 4D superrotation Ward identity is equivalent to a 2D conformal Ward identity.

The higher dimensional generalization of this result~\cite{Kapec2017}  has been worked out in~\cite{Kapec:2017gsg}.  Moreover, like the IR triangle this is only one of many iterations.  There is similarly a 2D Kac-Moody current corresponding to 4D large gauge transformations and the soft photon/gluon theorems~\cite{Strominger:2013lka,He:2015zea, Nande2018}.  The supertranslations also provide an abelian current~\cite{Strominger2014}.  The main impetus of this soft physics program~\cite{Strominger2017} is not just to understand the interconnectedness of IR universality and asymptotic symmetries on the 4D side.  Rather this story is intriguing to the next degree because of, and even began with, the realization that~\cite{Strominger:2013lka, Strominger2014} these symmetries appear to take the form of those of a 2D CFT.  And from the results explained in this section, we even have a notion of a stress tensor for this CFT.  However, if we hope to show that 4D scattering in asymptotically flat spacetimes admits a 2D holographic dual, we must know how to map more than just the currents.  We need to know how to construct the map~(\ref{eq:map42}).  Section~\ref{sec:cb} will take us beyond the IR to do just that.

\section{From 4D to 2D}\label{sec:cb}
In the previous section we showed that perturbative gravitational scattering in asymptotically flat spacetimes obeys an enhancement of the 4D global Lorentz algebra to superrotations.  These superrotations obey the same algebra as local conformal transformations of the celestial sphere -- a trait that seems like less of a coincidence when we consider the results of section~\ref{sec:stress}.  If there exists a 2D holographic dual, then this would just be the typical enhancement of global $SL(2,\mathbb{C})$ invariance to two copies of the Virasoro algebra~\cite{DiFrancesco:1997nk}.

The map we are after here is somewhat the reverse of a common trick in the CFT literature known as the embedding space formalism~\cite{Penedones:2016voo,SimmonsDuffin:2012uy,Costa:2011mg}.  It works for the same symmetry reason but the physics will be different.  In fitting with this program's revival and amalgamation of disparate formalisms developed in decades past, the embedding space formalism was introduced by Dirac in the 30s in \cite{10.2307/1968455}.  He even went on to study  `homogeneous expansor'  representations of the Lorentz group~\cite{doi:10.1098/rspa.1945.0003} which are curiously reminiscent of our conformal primary basis and dilation modes, but with a different spacetime-profile interpretation.

First, let us focus on the commonalities.  We introduce the embedding space following~\cite{SimmonsDuffin:2012uy}.  The $d$ dimensional Euclidean global conformal group is $SO(d+1,1)$, which is our beloved Lorentz group in $\mathbb{R}^{d+1,1}$ (now the `embedding space').  Moreover, the group action on the coordinates is linear in this space, but not on $\mathbb{R}^d$.  In particular the lightcone of the origin is mapped to itself.  Using $X^\mu$ to denote the coordinates of $\mathbb{R}^{d+1,1}$ and $x^a$ for $\mathbb{R}^d$, a $d$-dimensional Euclidean space can be formed from the quotient $X^\mu\sim \lambda X^\mu$, $\lambda\in\mathbb{R}$ within the null cone $X^2=0$, and the linear Lorentz transformations respect this scaling. 
 (For comparison our celestial sphere $\mathcal{CS}^2$ would be a quotient of null infinity by the generators of the lightcone of $i^0$ as in Figure~\ref{fig:penrose}.)  
    The Poincar\'e section is a gauge fixing of $\lambda$ that sets 
    \be\label{eq:projem}
 X^+=X^0+X^{d+1}=1,~~~X^-=X^0-X^{d+1}=x^2,~~~X^i=x^i~~\mathrm{for}~~i=\{1,...,d\}.
 \ee
To go from the linear action on $X$ to its image in the Poincar\'e section we have
 \be\label{eq:rescaling}
 X\mapsto \frac{\Lambda X}{(\Lambda X)^+}
 \ee
giving the standard non-linear action of conformal transformations on $\mathbb{R}^d$~\cite{SimmonsDuffin:2012uy}.  Moreover, primary fields are lifted to the lightcone of this embedding space.  A scalar primary of weight $\Delta$ is lifted to
\be\label{eq:homogeneousscale}
\Phi(X)=(X^+)^{-\Delta}\phi(X^\mu/X^+),
\ee
which then is just a scalar in $\mathbb{R}^{d+1,1}$.  Furthermore it is homogenous of degree $\Delta$ under a global dilation of the $X^\mu$.  This can be generalized to spinning fields where the embedding formalism helps in constructing fields with the appropriate transformation properties.  It was mentioned that this is somewhat the reverse of our map because correlators of $\Phi$ on embedding space are defined by the conformal correlators of $\phi$, while the symmetries of the former help determine the latter --  i.e.  conformal invariance of correlators in $\mathbb{R}^d$ is reduced to Lorentz invariance and homogeneity-matching of quantities constructed from vectors on the lightcone in~$\mathbb{R}^{d+1,1}$.  We will define CFT correlators in terms of $\mathbb{R}^{d+1,1}$ $\cs$-matrix elements, and we hope that once we understand the CFT side we can use that structure to learn something about scattering amplitudes in 4D.

In the end this trick is used to study a conformally invariant quantum theory in $d$ dimensions.  If this is supposed to be a Lorentzian conformal field theory, then an appropriate Wick rotation must be performed and the Hilbert space of this CFT should obey a unitary representation of the group $SO(d,2)$.  In this thesis we are concerned with a quantum theory in $\mathbb{R}^{d+1,1}$.  The Hilbert space will form a unitary representation of $SO(d+1,1)$ and any quantum field theory that we will end up coupling to gravity will have an $\cs$-matrix that obeys the usual locality and analyticity properties~\cite{Eden:1966dnq,Adams:2006sv,Bourjaily:2011cen}.  We are looking for a map that encodes 4D $\cs$-matrix elements into 2D CFT correlators, and it will be up to our future selves to figure out in what way the image of this map behaves like a `nice' CFT$_2$.

Now there are actually two natural $\mathbb{R}^{d+1,1}$ that we consider in ordinary quantum field theory:  position space, and momentum space.  We will find features very similar to the embedding space up-lift in both.  On the one hand, on-shell massless scattering amplitudes are naturally defined as functions on the lightcone in the  $\mathbb{R}^{d+1,1}$ momentum space~\cite{Pasterski:2017ylz}.  On the other hand, one can move off the lightcone in position space and consider a hyperbolic foliation, following de Boer and Solodukhin's early attempt at flat space holography~\cite{0303006dBS}.  This concept is to apply AdS/CFT (and dS/CFT) to each slice within (and outside) the forward/past lightcones of the origin.  Doing so muddles manifest translation invariance but emphasizes Lorentz covariance.  An interesting alternative is to consider not fixed $\tau^2=-X^2$, but rather fixed $\tau\p_\tau$ dilation eigenvalues.  Then free propagation of the Maxwell or linearized graviton field stays on this slice because the equations for different eigenvalues decouple.    
  This class of solutions contains a basis of the free wave equation that transform as conformal primaries~\cite{Pasterski:2017kqt}, which we will construct in section~\ref{sec:pbasis}.   The application of AdS/CFT tools and a momentum space picture combine in our analysis of the massive case.  Here the hyperboloid (or `Euclidean AdS') is not a slice of spacetime but rather the space of on-shell momenta $p^2=-m^2$.
  A similar construction has played a role in generalizing the Ward identity derivations to massive matter~\cite{Campiglia:2015qka,Campiglia:2015kxa, Campiglia:2015lxa}. These are related by the fact that near $i^\pm$, a massive particle's trajectory $ X^\mu(s)=\hat p^\mu s +X_0^\mu$ is dominated by $\hat{p}s$.
 
In section~\ref{sec:pb} we define conformal primary wavefunctions in position space and show that they form a basis for the single particle states when the dimensions lie on the principal series.  In section~\ref{sec:examples} we will provide some examples of amplitudes computed in this basis following \cite{Pasterski:2017kqt,Pasterski:2017ylz}.  

\subsection{A Conformal Primary Basis}\label{sec:pb}

Throughout this thesis our main focus has been on the 4D case relevant to our superrotation analysis (corresponding to $d=2$) for particular examples.  However since we have the scattering basis results worked out for arbitrary $d$, we will include them in section~\ref{sec:pbasis}. 

\subsubsection{Coordinate Conventions}\label{sec:conventions}
We begin by defining coordinates that will allow us to keep in closer touch with those used in the embedding formalism~(\ref{eq:projem}) without significantly modifying the story we have built thus far.  This will actually make some of the 2D expressions nicer.  Recall that Christoffel  and spin connection coefficients appeared in the stress tensor OPE~(\ref{eq:TOPE}).  These will go away if we go to coordinates with a flat celestial sphere metric, as in~\cite{Dumitrescu:2015fej}.
This amounts to replacing~(\ref{flt}) with
\be\begin{array}{rl}\label{eq:coo}
X^0=&\frac{1}{2}(u+r(1+z\bz))\\
X^1+iX^2=&r z\\
X^3=&\frac{1}{2}(-u+r(1-z\bz))\\
\end{array}\ee
for which the flat metric takes the form
\be\label{eq:flat}
ds^2=-dudr+r^2dzd\bz.
\ee
At leading order in large $r$ this amounts to the coordinate transformation
\be
u\mapsto\frac{u}{1+z\bz}+...,~~~r\mapsto \frac{r}{2}(1+z\bz)-\frac{u}{2}\frac{1-z\bz}{1+z\bz}+...,~~z\mapsto z+\frac{uz}{r(1+z\bz)}+...
\ee
starting from the round retarded radial coordinates of~(\ref{flt}).  Note that in terms of the embedding formalism~(\ref{eq:projem}), $X^+=X^0+X^3=r$.   In addition we have,
\be
X^\mu X_\mu=-ru
\ee and $u=0$ again describes the lightcone of the origin (recall the cross section~(\ref{eq:projem}) removes the point at infinity from the celestial sphere by assuming $X^+=r\neq0$, which is just one ray in~(\ref{eq:coo})).
Note that we can form the ratio
\be\label{eq:ratio}
z=\frac{X^1+iX^2}{X^0+X^3}.
\ee
If we remain on the lightcone, as in the embedding space formalism above, we find that Lorentz transformations act on this coordinate via a M\"obius transformation
\be\label{eq:sl2c1} z \to {az+b\over cz+d}\,\ee
where $\{a,b,c,d\}\in\mathbb{C}$ and $ad-bc=1$.  This is consistent since 4D Lorentz transformations are isomorphic to $SL(2,\mathbb{C})$. 
In particular, by forming the ratio~(\ref{eq:ratio}), the rescaling factors of $(\Lambda X)^+$ in~(\ref{eq:rescaling}) cancel.

For scattering in 4D spacetime, we need to consider what happens away from the lightcone, i.e. $u\neq0$.
By changing to coordinates
\be
u=\tau y,~~~r=\frac{\tau}{y}
\ee
one now has $X^\mu X_\mu=-\tau^2$ and the metric~(\ref{eq:flat}) becomes
\be\label{met}
ds^2=-d\tau^2+\tau^2 \left(\frac{dy^2+dzd\bz}{y^2}\right)
\ee
where the term in parenthesis is the metric on the hyperboloid $H_3$ in Poincar\'e coordinates $(y,z,\bz)$.
Here $\{\tau,y\}$ real and positive covers the forward lightcone, while $\tau\in\mathbb{R}_{<0}$ would cover the past light cone and $\tau\mapsto i\tau, y\mapsto iy$ covers the region outside the light cone. 
Considering bulk interactions slice by slice at each fixed $\tau$ is the starting point for the attempts at flat space holography by~\cite{0303006dBS} and more recently~\cite{Cheung:2016iub} in the context of the ongoing soft physics program.

When we study amplitudes in momentum space, replacing $X^\mu$ with $p^\mu$, energy positivity simplifies this geometry to only contain one region -- the forward lightcone.  Moreover, for single on-shell particle states, there is only one relevant surface:  the lightcone if the particle is massless, or the hyperboloid at fixed $p^2=-m^2$.  We can use the same Poincar\'e coordinates to parameterize a reference momentum on the unit mass hyperboloid
\begin{align}\label{map}
\hat p^\mu(y,z,\bar z) = \left({  1+y^2+|z|^2 \over 2y}  ,  {\text{Re} (z)\over y}, {\text{Im} (z)\over y}, {1-y^2-|z|^2\over 2y}    \right)\,.
\end{align}
The $SL(2,\mathbb{C})$ Lorentz action that transformed $z$ by a M\"obius transformation when on the lightcone now act as 
\begin{align}\label{sl2c}
&z\to z' = {(az+b)(\bar c\bar z +\bar d )+a\bar cy^2 \over|cz+d|^2+|c|^2y^2 } \,,\notag\\
&\bar z\to \bar z' = {(\bar a\bar z+\bar b)(c z + d )+\bar a cy^2 \over|cz+d|^2+| c|^2y^2 } \,,\notag\\
&y\to y' = {y\over |cz+d|^2+|c|^2y^2}\,,
\end{align}
on $H_3$.  As a crosscheck, one sees that as $y\rightarrow 0$ the transformation of $z$ approaches~(\ref{eq:sl2c1}) again, and the slices approach the boundary of the lightcone of the origin in this limit.  In our context it is thus natural to think of the celestial sphere where the putative 2D dual resides as being the boundary of the lightcone of the origin rather than an arbitrary cross section of $\mathcal{I}$ (or rather perhaps a separate past and future copy  $\mathcal{CS}^{2\pm}$ for in and out states~\cite{0303006dBS}).  This $u=0$ section is mapped to itself under Lorentz transformations, as well as their enhancement to superrotations, but not by translations. We thus do not expect translation invariance constraints to be as straightforward in the celestial sphere picture as they are in ordinary momentum space scattering amplitudes.

For future use we also introduce a massless reference vector for the metric~(\ref{eq:flat}) 
\begin{align}\label{eq:nullref}
q^\mu = \left( 1+|w|^2 , w+\bar w, -i (w-\bar w) , 1-|w|^2\right)\,
\end{align}
and point out that the scalar bulk to boundary propagator~\cite{Witten:1998qj} of dimension $\Delta$
\begin{align}\label{eq:gscalar}
G_\Delta (y, z,\bar z; w,\bar w) =  \left({ y\over y^2  +  |z-w|^2}\right)^\Delta\,
\end{align}
where $(y, z,\bar z)\in H_3$ and $(w,\bar w)\in\p H_3=\mathbb{C}$, and which transforms under $SL(2,\mathbb{C})$ as
\begin{align}\label{covariance}
&G_\Delta(y',z',\bar z'; w' ,\bar w') = |cw+d|^{2\Delta} G_\Delta(y,z,\bar z; w,\bar w)\,,
\end{align}
takes the simple form 
\be\label{eq:gsimple}
G_\Delta (y, z,\bar z; w,\bar w)=\frac{1}{(-\hat{p}\cdot q)^\Delta}
\ee
in terms of the $\mathbb{R}^{d+1,1}$ vectors $\hat{p}^\mu$ and $q^\mu$. Indeed most of our constructions in the following section will be inspired by embedding space formalism results.  

The coordinate conventions introduced thus far are relevant to the 4D scattering we have focused on in this thesis, where Lorentz transformations experience an enhancement to superrotations.  While we have a richer symmetry structure in the 4D/2D case, the generalization of the 2D stress tensor~\cite{Kapec2017} to arbitrary $d$~\cite{Kapec:2017gsg} indicates that we would also like to be able to prepare quantum field theory states ready to be coupled to gravity in this conformally covariant manner for $d\neq 2$.  In preparation for the generic-$d$ results quoted in the next subsection, we now describe how the above statements generalize.

The metric on $H_3$ that appears in~(\ref{met}) becomes 
\begin{align}
ds^2 _{H_{d+1}} =  {dy^2 +d \vec z \cdot d\vec z\over y^2}\,,
\end{align}
where as before $y>0$ but now $\vec{z}\in\mathbb{R}^d$ and we use an orthogonal metric on this Euclidean space when $d\neq2$.  The $SO(d+1,1)$ isometries act on these Poincar\'e coordinates as 
\begin{align}\label{isometry}
&\mathbb{R}^d~\text{translation}:~\,~~ y'  =y\,,~~~~~~~~\vec z\,'  = \vec z+\vec a\,,\notag\\
&SO(d) ~\text{rotation}:~ ~y' =y\,,~\,~~~~~~~ \vec z\,'= M\cdot \vec z\,,\\
&\text{dilation}:~~~~~~~~\,~~
~ y'= \lambda y\,,~~~~~~~\vec z\,'= \lambda \vec z\,, \notag\\
&\substack{\text{special conformal}\\\text{transformation}}:~~~~~
y'= {y\over 1+2 \vec b \cdot \vec z +|\vec b|^2 (y^2+|\vec z|^2)}\,,
~~~~
\vec z\,'  ={\vec z + (y^2 +|\vec z|^2) \vec b \over
 1+2 \vec b \cdot \vec z +|\vec b|^2 (y^2+|\vec z|^2)}\,.  \notag
\end{align} 
While taking the $y\rightarrow0$ limit, we have the coordinates of points on the boundary transforming in the standard non-linear way under conformal transformations
\begin{align}\label{CT}
&\mathbb{R}^d~\text{translation}:~~~~~~\vec w\,'  = \vec w+\vec a\,,\notag\\
&SO(d) ~\text{rotation}:~\,~~~ \vec w\,'= M\cdot \vec w\,,\\
&\text{dilation}:~~~~~\,~
~ ~~~~~~~\vec w\,'= \lambda \vec w\,, \notag\\
&\substack{\text{special conformal}\\\text{transformation}}:~~~~~~~~
\vec w\,'  ={\vec w + |\vec w|^2 \vec b \over
 1+2 \vec b \cdot \vec w +|\vec b|^2 |\vec w|^2}\,.  \notag
\end{align}
As in~(\ref{map}) and~(\ref{eq:nullref}), we can form unit mass and null reference momenta in terms of these bulk and boundary coordinates
\be\label{phat}
\hat p (y,\vec z) = \left( {1+y^2 + |\vec z|^2 \over 2y} \,,  \, { \vec z \over y} \,, \,{1-y^2 - |\vec z|^2 \over 2y}    \right),
\ee
which obeys $\hat{p}^2=-1$  and transforms linearly as 
\begin{align}\label{ptransform}
\hat p^\mu ( y' ,\vec z\,' )  = \Lambda^\mu_{~\nu} \hat p^\nu\,,
\end{align}
under~(\ref{isometry}) with $ \Lambda^\mu_{~\nu}$ the vector representation of the corresponding $SO(d+1,1)$ group element (with the representative chosen to be consistent with plugging~(\ref{isometry}) into~(\ref{phat}), respectively~(\ref{sl2c}) into~(\ref{map}) in the $d=2$ case).  Meanwhile the reference null direction
\begin{align}\label{qmap}
q^\mu (\vec w) =   \left( \, 1+ |\vec w|^2 \,,\,  2\vec w \, , \, 1-|\vec w|^2  \, \right),
\end{align}
maintains this form under the transformation 
\begin{align}\label{qtransform}
q^\mu (\vec w\,')  = \left| {\partial \vec w\,' \over \partial \vec w}\right|^{1/d} \, \Lambda^\mu_{~\nu} q^\nu(\vec w)\, ,
\end{align}
with the rescaling as in~(\ref{eq:rescaling}), familiar from our discussion of the selection of the lightcone section~(\ref{eq:projem}).  The scalar bulk-to-boundary propagator~(\ref{eq:gscalar}) of dimension $\Delta$ is now
\begin{align}\label{G}
G_{\Delta}(\hat p; \vec w)  = \left( {y\over y^2+ |\vec z- \vec w|^2}\right)^\Delta\,,
\end{align}
which like~(\ref{eq:gsimple}) can again be written in terms of~(\ref{phat}) and~(\ref{qmap})
\begin{align}\label{Gembedding}
G_\Delta( \hat p ;q )  = {1\over (-\hat p \cdot q)^\Delta}\,,
\end{align}
and as expected from~(\ref{qtransform})
\begin{align}\label{Gcovariance}
G_\Delta( \hat p '; q') =  \left| {\partial \vec w\,' \over \partial \vec w}\right|^{-\Delta/d}  \,G_\Delta(\hat p; q)\,.
\end{align}
As a final note, when we discuss spin we will use $\p_q q^\mu$ to denote a derivative with respect to $w^a$ which can be used to form the analog of the polarization tensors~(\ref{pol1}) with an $SO(d)$ vector index
\begin{align}\label{daq}
\p_a q^\mu \equiv {\p \over \p w^a }q^\mu(\vec w)  
= 2( w^a \,,\, \delta^{ba} \,,\, -w^a)\,.
\end{align}
We are now ready to describe our desired conformal basis.

\subsubsection{Conformal Primary Wavefunctions}\label{sec:pbasis}

In this section, we describe a basis of single particle scattering states for which $\mathcal{S}$-matrix elements in  $\mathbb{R}^{d+1,1}$ transform as $d$-dimensional conformal correlators under the action of $SO(d+1,1)$~\cite{Pasterski:2017kqt}.  We refer to the corresponding single particle modes as `conformal primary wavefunctions'.  These solutions to wave equation in $\mathbb{R}^{d+1,1}$ for the respective particle type (here we consider the Klein-Gordon, Maxwell, and linearized Einstein equations), are labeled by a point $\vec{w}\in\mathbb{R}^d$, a conformal dimension $\Delta$ -- which replace the usual parameterization of an on-shell momenta by direction and energy (or rapidity for the massive case) -- in addition to any appropriate $SO(d)$ spin indices.  We will find that the spectrum of conformal dimensions necessary to form a basis for the finite energy solutions lies in the principal series
\begin{align}\label{introprincipal}
\Delta\in \frac d2+ i \mathbb{R}\,,
\end{align}
independent of the mass of the field (unlike in AdS/CFT). (Note there is an interesting story still under development regarding certain zero modes in gauge theories which lie off the principal series but appear related to the asymptotic symmetry Goldstone modes and their conjugate soft modes~\cite{Donnay:2018neh}, see also section 4.1.1 of~\cite{Cheung:2016iub}).

We begin by quoting the relevant definitions introduced in~\cite{Pasterski:2017kqt}, and follow with a discussion of how to use results from section~\ref{sec:conventions} and embedding space formalism techniques from the CFT literature~\cite{10.2307/1968455,Mack:1969rr,Cornalba:2009ax,Weinberg:2010fx,Costa:2011mg,Costa:2011dw,SimmonsDuffin:2012uy,Costa:2014kfa} to identify the appropriate spectrum and verify completeness and orthonormality of this basis.  In each of the following the reference direction $\vec{w}$ transforms as in~(\ref{CT}) and $\Lambda^\mu_\nu$ refers to the representation~(\ref{ptransform}).
~\\

\noindent{\bf Definition 5.1} A \textit{scalar conformal primary wavefunction} $\phi_\Delta(X^\mu;\vec w)$ of mass $m$ is a wavefunction on $\mathbb{R}^{1,d+1}$ labeled by a ``conformal dimension" $\Delta$ and a point $\vec w$ in $\mathbb{R}^d$ which  satisfies the following properties
\begin{itemize}
\item It satisfies the $(d+2)$-dimensional  massive Klein-Gordon equation of mass $m$,
\begin{align}\label{KGsol}
\left( {\partial\over \partial X^\nu } {\partial\over \partial X_\nu} -m^2  \right) \phi_{\Delta}(X^\mu ;\vec w)=0\,.
\end{align}
\item It transforms covariantly as a scalar conformal primary operator in $d$ dimensions under an $SO(d+1,1)$ transformation,
\begin{align}\label{covariance}
\phi_\Delta \left( \Lambda^\mu_{~\nu} X^\nu ; \vec w\,'(\vec w) \right)
= \left| {\partial \vec w\,' \over\partial \vec w}\right|^{-\Delta/d} \,
\phi_\Delta( X^\mu ; \vec w)\,
\end{align}
where $\vec w\,'(\vec w)$ is an element of $SO(d+1,1)$ defined in \eqref{CT} and $\Lambda^\mu_{~\nu}$ is the associated group element in the $(d+2)$-dimensional representation.
\end{itemize}~~~~~~~~~~~~~~~~~~~~~~~~~~~~~~~~~~~~~~~~~~~~~~~

\noindent{\bf Definition 5.2}  A \textit{massless spin-one conformal primary wavefunction}  $A^{\Delta}_{\mu a}(X^\mu;\vec w)$  is a wavefunction on $\mathbb{R}^{1,d+1}$ labeled by a ``conformal dimension" $\Delta$, a point $\vec w$ in $\mathbb{R}^d$, an $SO(d+1,1)$ spacetime index $\mu$, and an $SO(d)$ spin index $a$  which  satisfies the following properties
\begin{itemize}
\item It satisfies the  $(d+2)$-dimensional Maxwell equation,
\begin{align}\label{eq:linee1}
\left(  {\partial\over \partial X^\sigma} {\partial\over \partial X_\sigma} \delta_{\nu }^\mu   - {\partial\over \partial X^\nu}{\partial\over \partial X_\mu}\right)    A^{\Delta }_{\mu a} (X^\rho;\vec w) =0\,.
\end{align}
\item It transforms both as a $(d+2)$-dimensional vector  and a $d$-dimensional spin-one conformal primary  with conformal dimension $\Delta$ under an $SO(d+1,1)$ Lorentz transformation
\begin{align}\label{covarianceV}
&A^{\Delta }_{\mu a } \left(\Lambda^\rho_{~\nu} X^\nu ; \vec w\,'(\vec w)\right)
=
{\partial {w}^b \over \partial {w'}^a}
\left| {\partial \vec w'\over \partial \vec w }\right|^{- (\Delta-1)/d}
\, \Lambda_\mu ^{~\sigma} A^{\Delta}_{\sigma b}(X^\rho;\vec w)\,,
\end{align}
where $\vec w\,'(\vec w)$ is an element of $SO(d+1,1)$ defined in \eqref{CT} and $\Lambda^\mu_{~\nu}$ is the associated group element in the $(d+2)$-dimensional representation.
\end{itemize}~~~~~~~~~~~~~~~~~~~~~~~~~~~~~~~~~~~~~~~~~~~~~~~

\noindent{\bf Definition 5.3} A \textit{massless spin-two conformal primary wavefunction} $h^{\Delta, \pm}_{\mu_1 \mu_2 ; a_1 a_2}(X^\mu; \vec w)$  is a wavefunction on $\mathbb{R}^{1,d+1}$ labeled by a ``conformal dimension" $\Delta$, a point $\vec w$ in $\mathbb{R}^d$, $SO(d+1,1)$ spacetime indices $\{\mu,\nu\}$, and $SO(d)$ spin indices $\{a_1,a_2\}$  which  satisfies the following properties
\begin{itemize}
\item It is symmetric both in the $(d+2)$- and $d$-dimensional vector indices and traceless in the latter
\begin{align}\label{trless}
\begin{split}
&h^{\Delta, \pm}_{\mu_1 \mu_2 ; a_1 a_2}=h^{\Delta, \pm}_{\mu_2 \mu_1 ; a_1 a_2}\,,~~~~
\\
&h^{\Delta, \pm}_{\mu_1 \mu_2 ; a_1 a_2}=h^{\Delta, \pm}_{\mu_1 \mu_2 ; a_2 a_1}\,,~~~~
\delta^{a_1a_2}h^{\Delta, \pm}_{\mu_1 \mu_2 ; a_1 a_2}=0\,.
\end{split}
\end{align}
\item It is a solution to the vacuum linearized Einstein equations in flat space
\begin{align}\label{linee}
 \p_\sigma \p_\nu h^{\sigma}_{~\mu ;a_1a_2}
+\p_\sigma \p_\mu h^\sigma_{~\nu;a_1a_2}
-\p_\mu \p_\nu h^{\sigma}_{~\sigma;a_1a_2}
-\p^\rho \p_\rho h_{\mu\nu;a_1a_2}
=0\,.
\end{align}
\item It transforms both as a $(d+2)$-dimensional rank-two tensor  and a $d$-dimensional spin-two conformal primary  with conformal dimension $\Delta$ under an $SO(d+1,1)$ Lorentz transformation
\begin{align}
\label{covarianceh}
&h^{\Delta, \pm}_{\mu_1\mu_2; a_1a_2 } \left(\Lambda^\rho_{~\nu} X^\nu ; \vec w\,'(\vec w)\right)
=\scalemath{0.8}{
{\partial {w}^{b_1} \over \partial {w'}^{a_1}}
{\partial {w}^{b_2} \over \partial {w'}^{a_2}}
\left| {\partial \vec w'\over \partial \vec w }\right|^{- (\Delta-2)/d}
\, \Lambda_{\mu_1} ^{~\sigma_1}\, \Lambda_{\mu_2} ^{~\sigma_2}}
 h^{\Delta,\pm}_{\sigma_1\sigma_2 ;  b_1b_2}(X^\rho;\vec w)\,,
\end{align}
where $\vec w\,'(\vec w)$ is an element of $SO(d+1,1)$ defined in \eqref{CT} and $\Lambda^\mu_{~\nu}$ is the associated group element in the $(d+2)$-dimensional representation.
\end{itemize}

As long as we can show that they have finite norm, these solutions can be expanded in the standard plane wave single particle wavefunctions since they also satisfy the respective equations of motion and form a complete basis.  We are thus after functions supported on the on-shell hyperboloid/light cone in the massless cases, such that their Fourier transforms to position space satisfy the requirements of the respective Definitions 5.1-5.3.

The transformation properties of the bulk to boundary propagator~(\ref{Gcovariance}) discussed in section~\ref{sec:conventions} are enough to write down an ansatz for the massive scalar case which will satisfy Definition 5.1
\begin{align}\label{CPW}
\,
\phi^\pm_\Delta(X^\mu ;  \vec w) 
= \int_{H_{d+1}} [d\hat p ]  \, G_\Delta ( \hat p ;\vec w)  \, \exp \left[
\, \pm i m \hat p \cdot X \,\right]
\,,
\end{align}
where $ [d\hat p ]$ is the appropriate Lorentz invariant measure on $H_{d+1}$:
\begin{align}
\int_{H_{d+1}} [d \hat p] \equiv \int_0^\infty {dy\over y^{d+1} } \int d^d\vec z  =  \int {d^{d+1} \hat p^i \over \hat p^0 } \,
\end{align}
where $i=1,2,\cdots,d+1$ and $\hat p^0 = \sqrt{ \hat p^i \hat p^i +1}$.  The $d=2$ analog of this using~(\ref{covariance}) was noticed in~\cite{Pasterski:2016qvg}.  The position space profile can be explicitly evaluated in terms of a modified Bessel function of the second kind with finite Klein Gordon norm.  What is relevant to us here is that an $i\e$ prescription is needed to avoid the singularity when $q\cdot X=0$~\cite{Pasterski:2017kqt}, with a sign that depends on whether the particle is incoming or outgoing (see also~\cite{Donnay:2018neh} for an example where how one avoids the additional singularity at the lightcone $X^2=0$ is relevant).  Note that all of the singularities are outside of the Milne region where  the contributions in the spacetime slicing approach are most tractable and also more in common with standard AdS/CFT Witten-diagram techniques~\cite{Cheung:2016iub}.  This, combined with the simplicity of~(\ref{CPW}), shows that there is much to be gained by avoiding this slicing and going directly to a transform on the momentum space amplitudes as in~\cite{Pasterski:2016qvg,Pasterski:2017ylz}.  Indeed acting on the scalar $\cs$-matrix element $\mathcal{A} ( \pm m_i \hat p_i^\mu)$
\begin{align}\label{integral1}
\mathcal{\hat A}(\Delta_i , \vec w_i )  
\equiv \prod_{k=1}^n \int_{H_{d+1}} [d\hat p_k] \, 
G_{\Delta_k} (\hat p _k ; \vec w_k )  \, \
\mathcal{A} ( \pm m_i \hat p_i^\mu)\,,
\end{align}
we find that
\begin{align}\label{conformalcorrelator}
\mathcal{\hat A}(\Delta_i, \vec w_i'(\vec w_i) ) 
= \prod_{k=1}^n \left|  {\partial \vec w_k' \over \partial \vec w_k}\right|^{-\Delta_k/d}\,
\mathcal{\hat A}(\Delta_i,\vec w_i ) \,
\end{align}
which transforms like a $d$-dimensional conformal correlator.  We have constructed our first example of the map~(\ref{eq:map42}) (with the same result for the particular case of $d=2$ appearing in~\cite{Pasterski:2016qvg}).  We note that in contrast to the notation in~\cite{Pasterski:2016qvg, Pasterski:2017kqt, Pasterski:2017ylz}, we will use a carat to denote amplitudes transformed under~(\ref{eq:map42}) and a tilde to denote conformal shadows.

Now that we have a profile currently defined, at least up to analytic continuation, for arbitrary $\Delta$, we want to determine what spectrum is needed to form a complete basis (and not over-complete).  For this we will use tricks from the CFT literature.  For example, the shadow transformation maps a spin $J$ conformal (quasi)-primary wavefunction of dimension $\Delta$ to one of dimensions $d-\Delta$ (we restrict to integer $J$ here and note that in $d=2$ the helicity flips under this transformation~\cite{Dolan:2011dv,Osborn:2012vt}) and amounts to an integral transform on $\mathbb{R}^{d}$.  From the point of our basis quest, this amounts to a linear relation.  In particular the shadow of $\phi^\pm_\Delta$ in (\ref{CPW}) is $\phi^\pm_{d-\Delta}$.  Thus, if these ranges overlap as they do for the principal series~(\ref{introprincipal}), half of this spectrum will be redundant.

Using results from the CFT literature~\cite{Costa:2014kfa}, we have orthonormality relations for the bulk-to-boundary propagators~(\ref{G})  when the weights lie on the principal series~(\ref{introprincipal}).  First, on the space $\{\Delta,\vec{w}\}$ that labels the conformal primary wavefunctions satisfying Definition 5.1 we have
 \begin{align}\label{ortho1}
 \int_{-\infty}^\infty d\nu \,
\mu(\nu)
 \,  \int d^d\vec w \, G_{\frac d2+i\nu}(\hat p_1; \vec w)G_{\frac d2-i\nu}(\hat p_2;\vec w)   =
   \, \delta^{(d+1)}(\hat p_1,\hat p_2)\,,
 \end{align}
where the measure $\mu$ takes the form 
 \begin{align}
 \mu(\nu)   = {\Gamma(\frac d2 +i\nu  )\Gamma(\frac d2 -i\nu) \over
  4\pi^{d+1}  \Gamma(i\nu)\Gamma(-i\nu)}\,,
 \end{align}
and also on the kinematic space $\hat{p}$
 \begin{align}\label{ortho2}
& \int_{H_{d+1}}[d\hat p]\, G_{\frac d2+i\nu}(\hat p;\vec w_1) 
 G_{\frac d 2+i \bar\nu}(\hat p;\vec w_2)
 =\\
&2\pi^{d+1} 
{\Gamma(i\nu)\Gamma(-i\nu) \over \Gamma(\frac d2 +i\nu) \Gamma(\frac d2 -i\nu)}
 \delta(\nu+\bar \nu) 
 \delta^{(d)}(\vec w_1-\vec w_2)
 + 2\pi^{\frac d2+1} 
   {\Gamma(i\nu)\over \Gamma(\frac d2+i\nu)}
     {\delta(\nu-\bar\nu)}
   {1\over |\vec w_1-\vec w_2|^{2(\frac d2+i\nu)}}\,.\notag
 \end{align}
We see that this indeed reduces to an orthonormality condition when we
note that, because $d-\Delta=\Delta^*$ on the principal series, the complete basis is spanned by 
\be\label{eq:spec}
\Delta\in\frac{d}{2}+i\mathbb{R}_{\ge0}.
\ee
Indeed we use these equations to invert~(\ref{CPW}) and expand the ordinary plane waves in terms of our conformal primary wavefunctions with the spectrum~(\ref{eq:spec})
\begin{align}\label{inverse}
e^{\pm im \hat p\cdot X}  =2\int_{0}^\infty d\nu \,\mu(\nu) \int d^d\vec w \,\,
G_{\frac d2-i\nu}(\hat p;\vec w) \,\, \phi^\pm_{\frac d2+i\nu}(X^\mu;\vec w)  \,.
\end{align}
One can further show that the above integrals imply that the Klein-Gordon inner product 
\begin{align}\label{KGip}
( \Phi_1 , \Phi_2)&=- i \int d^{d+1}X^i~\left[\,  
\Phi_{1} ( X) \, \partial_{X^0}\Phi_{2}^{*} ( X)
-\partial_{X^0}\Phi_{1} ( X) \,\Phi_{2}^{*} ( X)
\right] \,,
\end{align}
on the $\phi^+_{\frac{d}{2}+i\nu}$ is positive and proportional to $\delta(\nu_1- \nu_2) \, \delta^{(d)}(\vec w_1- \vec w_2)$.  We thus have verified the completeness and orthonormality properties of our conformal basis for a massive scalar.  Note that this inner product and the analog of~(\ref{ortho2}) would be divergent if we were off the principal series.

We can take the massless limit of the above construction by defining $\omega\equiv\frac{m}{2y}$ and using the $y\rightarrow0$ boundary behavior of the bulk to boundary propagator
\begin{align}\label{bdylimit}
G_\Delta(y,\vec z;\vec w)
~\underset{m\to0}{\longrightarrow }~
\pi^{d\over2}   {\Gamma(\Delta-\frac d2)\over\Gamma(\Delta)} 
y^{d-\Delta} \delta^{(d)}(\vec z-\vec w)
+{y^\Delta\over |\vec z-\vec w|^{2\Delta}}+\cdots\,
\end{align}
When on the principal series both terms have an absolute value that scales with the same (non-zero) power of $y$, and since $y=\frac{m}{2\omega}$, the differing phases make the $m\rightarrow0$ limit ill defined.  However, both terms separately satisfy Definition 5.1 and one is the conformal shadow of the other.  We thus take our massless scalar conformal primary to be the Mellin transform
\begin{align}\label{m0CPW} 
 \varphi^\pm _{\Delta} (X^\mu; \vec w)\equiv
\int_0^\infty d\omega \,\omega^{\Delta-1 }\,
e^{\pm i \omega q \cdot X-\epsilon \omega}
={ (\mp i)^\Delta \Gamma(\Delta)  \over  (-q(\vec w)\cdot X\mp i\epsilon)^{\Delta}}  \,,
\end{align}
and because we no longer have the shadow redundancy, we need the full principal series to form our basis
\be\label{eq:spec2}
\Delta\in\frac{d}{2}+i\mathbb{R}.
\ee
Curiously these $\{\Delta,\vec{w}\}$ provide a $2:1$ map of massless to massive solutions of the respective Klein-Gordon equations.  This is a na\"ive doubling of the size of solution space.  In the standard momentum space picture this would be like trying to compare the `size' of the lightcone to the fixed $m^2$ hyperboloid.  Projecting down to the space of three momenta would give a $1:1$ map between these spaces.  

Using the Mellin and inverse Mellin transform definitions
\begin{align}
{\hat f}(\Delta) = \int_0^\infty d\omega \, \omega^{\Delta-1} f(\omega)\,,
\end{align}
and for its inverse
\begin{align}\label{Mellin}
f(\omega) = {1\over 2\pi i} \int_{c-i\infty}^{c+i\infty} d\Delta \, \omega^{-\Delta} \hat f(\Delta)\,,~~~~~c\in \mathbb{R}\,.
\end{align}
one can invert~(\ref{m0CPW}) to expand the massless planes waves on~(\ref{eq:spec2}), and using the identity
\begin{align}\label{m02pt}
\int_0^\infty d\omega \, \omega^{i \nu -1} 
= 2\pi \delta(\nu)\,
\end{align}
one can show that these are orthonormal under the Klein-Gordon inner product.  Thus we again have established that these form a basis.

We see that the integral transform to go to the conformal correlators is simpler in the massless case than the massive one since these particles already have a natural reference direction on the celestial sphere.  We only need to integrate over the null ray in momentum space with the appropriate weight 
\begin{align}\label{integral}
\mathcal{\hat A}(\Delta_i , \vec w_i )  
\equiv \prod_{i=1}^n  \int_0^\infty d\omega_i  \, \omega_i^{\Delta_i-1 }  \, \
\mathcal{A} ( \pm \omega_i \hat q_i^\mu)\,.
\end{align}
The Mellin solutions~(\ref{m0CPW}) were studied in~\cite{0303006dBS,Cheung:2016iub}. Note that the final form of~(\ref{m0CPW}) as compared to the bulk-to-boundary propagator~(\ref{G}) extends the AdS result to arbitrary slices while also satisfying the homogeneity property~(\ref{eq:homogeneousscale}) that embedding space fields would obey (recalling those would be defined on the lightcone rather than the unit hyperboloid).  

Moving on to Definition 5.2 for the Maxwell field, as in~\cite{Cheung:2016iub} we can the spin-one bulk-to-boundary propagator to construct our wavefunction.  The propagator is constructed using the embedding space formalism where the polarization tensors~(\ref{daq}) are used to project down to $\mathbb{R}^d$ for the spin index.  We quote the resulting wavefunction~\cite{Pasterski:2017kqt}
 \begin{align}\label{spin1cpw} 
A_{\mu a}^{\Delta, \pm } (X^\mu ; \vec w)  =-{1\over (-q\cdot X\mp i\epsilon)^{\Delta-1}}
{\p \over \p X^\mu } {\p \over \p w^a } 
\log (-q\cdot X\mp i\epsilon).\,
\end{align}
The conformal shadow acts simply
\begin{align}\label{spin1shadow}
\widetilde{A_{\mu a}^{\Delta, \pm }} (X^\mu ; \vec w)  = (-X^2)^{\frac d2-\Delta} A_{\mu a}^{d-\Delta, \pm } (X^\mu ; \vec w) \,,
\end{align}
and this solution again satisfies Definition 5.2.  By definition of the shadow these solutions are linear combinations of the~(\ref{spin1cpw}) and since they have a different functional form there is no redundancy we again need the full principal series spectrum.  The conformal primary solutions~(\ref{spin1cpw}) and their shadows~(\ref{spin1shadow}) obey both the radial and harmonic gauge conditions
\begin{align}\label{eq:gauge}
X^\mu A_{\mu a}^{\Delta, \pm }(X^\mu ; \vec w)=0\,,~~~~~ \p^\mu A_{\mu a}^{\Delta, \pm }(X^\mu ; \vec w)=0\,
\end{align}
which are mutually compatible for solutions of the free Maxwell equation~\cite{Magliaro:2007qr} but not in the presence of sources.  We thus find that demanding conformal covariance of the wavefunction fixes the gauge.  Of course we expect the scattering amplitudes to be gauge invariant.

Note that up to an overall factor, the solutions~(\ref{spin1cpw}) are gauge equivalent to a Mellin transform of $\p_a q_\mu 
e^{\pm i \omega q\cdot X-\epsilon \omega}$.  Moreover they differ by a gauge transformation that breaks the radial condition in~(\ref{eq:gauge}), but preserves the Lorenz condition.  The spin-one analog of the Klein-Gordon inner product for scalars~(\ref{KGip}) is now
\cite{Ash3,Crnkovic:1986ex,Lee:1990nz,Wald:1999wa}:
 \begin{align}\label{spin1ip}
 (A_\mu , A'_{\mu'}) =- i \int d^{d+1}X^i \,\left[ A^{\rho} {F
'}_{0\rho}^{*} 
 -{A'}^{ \rho *} F_{0\rho}
 \right] \,.
 \end{align}
 By the Maxwell equations, this inner product is invariant under changing the Cauchy slice and gauge transformations as long as fields falloff rapidly at the boundary.  Boundary terms do arise for large gauge transformations which have a non-trivial symplectic pairing with the physical soft modes discussed above in sections~\ref{sec:soft} and~\ref{sec:mem}.  As long as we restrict our considerations to strictly non-zero energy radiative modes, the relation~(\ref{m02pt}) applied to a Mellin transform of the plane wave inner product is enough to make our basis claims and satisfy Definition 5.2.  We can study amplitudes in this basis by performing the transform~(\ref{integral}) with $\Delta$ on the principal series,  providing the desired map~(\ref{eq:map42}) for gauge fields.
 
A very similar story applies to the gravitational case satisfying Definition 5.3.  We find the conformal primary wavefunctions
\be\begin{array}{rl}\label{spin2cpw}
h^{\Delta,\pm}_{\mu_1\mu_2 ;a_1a_2}(X; \vec w)
&=P^{b_1b_2}_{a_1a_2}\,
{1\over (-q\cdot X\mp i\epsilon)^{\Delta-2}}
\partial_{b_1}\partial_{\mu_1}
\log (-q\cdot X\mp i\epsilon) \,\,
\partial_{b_2}\partial_{\mu_2}
\log (-q\cdot X\mp i\epsilon) \,,
\end{array}\ee
where we have employed the traceless symmetric projector
\begin{align}
P^{b_1b_2}_{a_1a_2}\equiv 
\delta^{b_1}_{~(a_1} \delta^{b_2}_{~a_2)} - \frac 1d \delta_{a_1a_2}\delta^{b_1b_2}\,,
\end{align}
 to satisfy condition~(\ref{trless}). As for the spin-one case~(\ref{spin1shadow})
the conformal shadow again takes a simple form
\begin{align}\label{spin2shadow}
\widetilde{h^{\Delta,\pm}_{\mu_1\mu_2 ;a_1a_2}}(X; \vec w)
=(-X^2)^{\frac d2 -\Delta }h^{d-\Delta,\pm}_{\mu_1\mu_2 ;a_1a_2}(X; \vec w)\,.
\end{align}
Both~(\ref{spin2cpw}) and the shadow~(\ref{spin2shadow}) satisify the radial and harmonic gauge conditions, in addition to being traceless on the $SO(d+1,1)$ indices
\begin{align}\label{fixdiff}
\eta^{\mu_1\mu_2}h^{\Delta,\pm}_{ \mu_1\mu_2; a_1a_2}
=0\,,~~~~~~\partial^{\mu} h^{\Delta,\pm}_{ \mu\mu_2; a_1a_2}
=0\,,~~~~~~X^\mu h^{\Delta , \pm}_{ \mu\mu_2; a_1a_2}=0\,
\end{align}
which are compatible when there are no stress tensor sources~\cite{Magliaro:2007qr}.  With this gauge fixing the wave equation~(\ref{linee}) reduces to $\Box h_{\mu\nu}=0$.  As in the spin-one case, the solutions~(\ref{spin2cpw}) are gauge equivalent to a Mellin transform of the ordinary plane wave solutions $P^{b_1b_2}_{a_1a_2} \,\p_{b_1} q_\mu \p _{b_2} q_\nu \, e^{\pm i\omega q\cdot X}$, and one again can show they form a basis with $\Delta$ on the principal series by applying a Mellin transform to the inner product
\begin{align}\label{spin2ip}
\left( h_{\mu\nu} , h'_{\mu'\nu'}\right)
&= -{ i} \int d^{d+1}X^i \Big[\,
h^{\mu\nu} \p_0 {h'}^*_{\mu \nu}  - 2h^{\mu\nu} \p_\mu {h'}^*_{0\nu}
+h\p^\mu {h'}^*_{0\mu } - h\p_0 {h'}^*+ h_{ 0\mu}\p^\mu {h'}^*  \notag\\
&~~~ -(h\leftrightarrow {h'}^*)\,
\Big]\,,
\end{align}
evaluated on the plane wave basis. 

So far in this section we have avoided talking about the zero frequency modes which have been a staple most of what we had been doing before this point.  There is a sense in which the $\omega\neq0$ modes were all that was missing from the map~(\ref{eq:map42}).  Indeed, the realization that soft modes mapped to currents is what started this program~\cite{Strominger:2013lka,Strominger2014} and inspired us to look for this map.  However, understanding how these $\omega\rightarrow0$ modes translate to our conformal primary basis is important if this basis indeed is the preferred description for understanding a holographic dual.  As pointed out in~\cite{Cheung:2016iub}, the ordinary soft limit contributes to the the `conformally soft' limit of $\lambda\rightarrow0$ on the principal series.  Of the conformal primary wave functions we have constructed above, in addition to the principal series which captures the radiative finite energy modes, there are some other $\Delta$ for which (\ref{spin1cpw}), (\ref{spin1shadow}), (\ref{spin2cpw}), or (\ref{spin2shadow}) become pure gauge, as summarized in Tables~\ref{table:puregauge} and~\ref{table:purediff}.  Some of the entries of this table have appeared in the analyses~\cite{Cheung:2016iub, Donnay:2018neh} and work to understand their role as well as non-principal series values of $\Delta$ is ongoing.
~\\

 \begin{table}[h!]
\centering
\caption{Spin-one conformal primary wavefunctions which are pure gauge in various dimensions.  For $d=2$, the conformal primary  wavefunction  $A^{\Delta=1,\pm} _{\mu a}$ is its own shadow.}\label{table:puregauge}
\begin{tabular}{|c|c|c|c|}
\hline
 & ~~~~$d=2$~ ~~~
&~~~ $d\neq 2$~ ~~\\
\hline
&&\\
~~~$A_{\mu a}^{\Delta,\pm}$~~~&  
&$\Delta=1$ \\
&$~  \Delta=1~$&\\
$\widetilde{A_{\mu a}^{d-\Delta,\pm}}$ & &$ \times$\\
&&\\
\hline
\end{tabular}
\end{table}

 \begin{table}[h!]
\centering\caption{Spin-two conformal primary wavefunctions which are pure gauge in various dimensions.  For $d=2$, the conformal primary  wavefunction  $h^{\Delta=1,\pm} _{\mu_1\mu_2 ;a_1a_2}$ is its own shadow.}
\label{table:purediff}
\begin{tabular}{|c|cc|c|c|}
\hline
 & ~~~~~~~~~~~~~~~~$d=2$&
&~~~ $d\neq 2$~ ~~\\
\hline
&&&\\
~~~$h^{\Delta,\pm} _{\mu_1\mu_2 ;a_1a_2}$~~~&  
&~$\Delta=0$~~~&$\Delta=0,1$ \\
&$ \Delta=1$&&\\
$\widetilde{ h^{d-\Delta,\pm} _{\mu_1\mu_2 ;a_1a_2}}$& &\, $\Delta=2$~~~&$ \times$\\
&&&\\
\hline
\end{tabular}
\end{table}
~\\

For example, while the conformal primary transformation conditions~(\ref{covarianceV}) and~(\ref{covarianceh}) select solutions which obey the gauge conditions~(\ref{eq:gauge}) and~(\ref{fixdiff}), it is clear that the converse is not true.  The fact that we can form a basis of free solutions implies we can smear such solutions on the celestial sphere to get the most general current-free case.  However, one might ask what happens when we have matter sources, especially considering that our soft theorem/memory relations in section~\ref{sec:mem} relied on an application of constraint equations with such charge and stress tensor sources. We note that the seemingly coincidental (from our formulation, not once one decomposes the massless wave equation taking into account $SL(2,\mathbb{C})$ Casimirs), homogeneity under  $X\rightarrow \lambda X$ of the massless solutions
~(\ref{m0CPW}), (\ref{spin1cpw}),  (\ref{spin2cpw}) and their shadows resembles that of the embedding space formalism~(\ref{eq:homogeneousscale})  (taking position space as an extension of the embedding space beyond the lightcone).  

This is a symptom of a more general feature that the linearized wave equations~(\ref{linee}),~(\ref{eq:linee1}), and the massless version of~(\ref{KGsol}) decouple for different values of the dilation eigenvalue $\tau\p_\tau$, and solutions with sources can be handled in a similar way.  The equations at fixed $\Delta$ then take the form of  AdS or dS wave equations which can be solved in a boundary to bulk manner.  We see, as in~\cite{0303006dBS}, how $d+2$ dimensional scattering data is encoded in data at the celestial spheres $\mathcal{CS}^{d+}$ and $\mathcal{CS}^{d-}$.  Moreover, because these equations are Lorentz invariant (so we can go to the rest frame) and linear (we can superimpose solutions), and we know from electromagnetism~\cite{Jackson:1998nia} that higher multipole configurations fall off at faster powers of $r$, we would expect a tower of integer $\Delta$ modes for the gauge field that would not contribute to the radiative states but should be an important part of understanding the holographic dual of these states.  The proposed infinite towers of gravitational memories suggested in~\cite{Compere:2019odm} computed in harmonic gauge near null infinity seem like they would gel with this picture.  Thus this avenue is worth pursuing further.  

We conclude this section by noting that the construction of the conformal basis informs our understanding of the soft sector beyond the questions it raises about how to handle the large gauge modes.  In particular, the discussion of conformal shadows, which appeared here in the context of trying to insure our basis was not over-complete also ties into our soft physics story.  When forming a conformal primary scattering basis one has a choice of whether or not to shadow or form some linear combination and still get a quasi-primary transformation law.  However, it was noticed in~\cite{He2014}  (before magnetic contributions were incorporated as a modification of Weinberg's soft photon theorem~(\ref{eq:soft}) in~\cite{Strominger2016b}) that the combination
\begin{equation}
{\bf a}_-\equiv a_-(\omega\hat{x})-\frac{1}{2\pi}\int d^2w\frac{1}{\bz-\bw}\p_\bw a_+(\omega\hat{y})
\end{equation}
has no pole as $\omega\rightarrow0$ a result of a relation between the soft photon theorem for $\pm$ helicities.  In the language of shadows, this says that a certain combination of Mellin and shadow of the opposite helicity decouples in the conformally soft limit.  This interpretation has helped elucidate certain independent logarithmic modes that appear in the $\lambda\rightarrow0$ limit in~\cite{Donnay:2018neh}.  It also provides a cleaner interpretation of the stress tensor in~\cite{Kapec:2017gsg}.

\vspace{2em}

\subsection{Some Examples}\label{sec:examples}
Now that we have constructed a conformal primary basis for various particle types, we return to the realm of 4D scattering amplitudes and apply the maps and~(\ref{integral1}) and~(\ref{integral}) to familiar momentum space amplitudes to see what happens.  Our amplitude transforms were designed to ensure the result transforms covariantly as 2D conformal correlators of quasi-primary operators corresponding to each external state.
~\\

We begin with a toy example from~\cite{Pasterski:2016qvg} of a `near-extremal' massive cubic decay.  Consider a local cubic interaction
\be
  \mathcal{L} \sim {\lambda } \phi_1 \phi_2  \phi_3+\cdots\,
\ee
so that the momentum space amplitude is a delta function for the sum of momenta
\be 
\mathcal{A}(p_i)= i(2\pi)^4\lambda\, \delta^{(4)}(- p_1+p_2+p_3)\,.
\ee
If the $\phi_i$ have arbitrary masses $m_i$, then we are integrating with bulk to boundary propagators over three different hyperboloids in momentum space, with the constraint that any configuration contributing to this integral obeys the above momentum conservation condition.  We see that our transform is probing the structure of kinematically allowed configurations. 

The fact that our transform is designed to make the result a conformally covariant `correlator' of quasi-primaries guarantees that the final amplitude should have the form
\be \mathcal{ \hat A}(w_i,\bar w_i)\sim {\lambda \over |w_1-w_2|^{\Delta_1+\Delta_2-\Delta_3}|w_2-w_3|^{\Delta_2+\Delta_3-\Delta_1}|w_3-w_1|^{\Delta_3+\Delta_1-\Delta_2}}\,,\ee
as long as the integral converges and does not have some more singular support (we will see examples in the massless case where we can solve the equations imposed by 2D conformal symmetry~\cite{DiFrancesco:1997nk} with delta function supported terms).  For practice, we will pick a mass configuration that makes the integrals much simpler to evaluate:  letting $m_2=m_3=m$ and $m_1=2(1+\e)m$ with $\e\rightarrow0$ what we will call a near extremal limit.  This is relevant for describing a decay process where a composite particle splits into daughter particles where the binding energy was small, so that they emerge with only small kinetic energies in the center of mass frame (a similar setup where the daughter particles have unequal masses should share many of the simplifications we find). Using the 4D analog of~(\ref{integral1}), we must now perform the integral
\begin{align}\label{eq:eval}
\mathcal{ \hat A}(w_i,\bar w_i) &=i(2\pi)^4 \lambda m^{-4}
\left( \prod_{i=1}^3 \int_0^\infty   {dy_i \over y_i^3} \int dz_id\bar z_i \, \right)\\
&~~~~~~~~~~~
\times\prod_{i=1}^3 G_{\Delta_i} (y_i , z_i,\bar z_i ;  w_i,\bar w_i)\,
\delta^{(4)}(-2(1+\e) \hat p_1 +\hat p_2+\hat p_3)\,\notag,
\end{align}
with $\hat p^\mu (y_i , z_i,\bar z_i)$ as in~(\ref{map}).  One of the $(y_i,z_i,\bz_i)$ integrals is saturated by part of the delta function, leaving a single remaining delta function constraint within a six-dimensional integral, so that we have five independent integration variables remaining.

In the~$\e\rightarrow0$ limit the three momenta become collinear so that the momentum transform begins to resemble that of an AdS/CFT Witten-diagram with a local interaction propagated to three boundary points labeled by the $(z_i,\bz_i)$.  By series expanding in small $\e$ we get terms proportional to a three dimensional $(y,z,\bz)$ integral of this form, with the remaining two dimensional integral doable and finite.  After the appropriate change of variables and integrations are performed, we find that~(\ref{eq:eval}) evaluates to  
\be\begin{array}{c}\label{main}
\scalemath{0.9}{\mathcal{ \hat A}(w_i,\bar w_i) 
=}\scalemath{0.9}{\frac{C(\Delta_i)}{ |w_1-w_2|^{\Delta_1+\Delta_2-\Delta_3}|w_2-w_3|^{\Delta_2+\Delta_3-\Delta_1}|w_3-w_1|^{\Delta_3+\Delta_1-\Delta_2}} }+\mathcal{O}(\e)\,
\end{array}\ee
where
\be
C(\Delta_i)={ i 2^{9\over2}\pi^6\lambda \Gamma({\Delta_1+\Delta_2+\Delta_3-2\over 2}) \Gamma({\Delta_1+\Delta_2-\Delta_3\over 2})\Gamma({\Delta_1-\Delta_2+\Delta_3\over 2})\Gamma({-\Delta_1+\Delta_2+\Delta_3\over 2})\sqrt{\e}\over m^4
\Gamma(\Delta_1)\Gamma(\Delta_2)\Gamma(\Delta_3)},
\ee
and from our discussion in section~\ref{sec:pbasis}, the weights should lie on the principal series $\Delta_j\in 1+i\lambda_j$.  We see from this result that we indeed get the expected form, but that most of the features of the final result come from the kinematics of the scattering.  We will see this has more dramatic effects for transforms of massless amplitudes, which we turn to next.

Following~\cite{Pasterski:2017ylz}, the transform we now wish to apply is~(\ref{integral}) with $\Delta_i=1+i\lambda_i$, and we will be considering color stripped gluon amplitudes here.  Since the equations for arbitrary $d$ above do not highlight the simplification when $d=2$, we first note that equation~(\ref{covarianceV}) reduces to
\begin{align}\label{covarianceV2}
&A^{\Delta}_{\mu J } \left(\Lambda^\mu_{~\nu} X^\nu ; {az+b\over cz+d} , {\bar a \bar z + \bar b \over \bar c \bar z +\bar d}\right)
= (cz+d)^{\Delta+ J }  (\bar c\bar z +\bar d)^{\Delta- J}\, \Lambda_\mu ^{~\rho} A^{\Delta}_{\rho J}(X^\mu;z,\bar z)\,.
\end{align}
We further note that outgoing one-particle states with 4D helicity $\ell=\pm1$  correspond to 2D operators with spin $J=\pm1$, while for incoming states the sign is reversed.  We will label amplitudes with the helicity the particles would have if they were all outgoing, though we will need to keep track of which are incoming and which are outgoing because as we saw in the massive example above, our integral transform acts on the momentum conserving delta function.  Because each particle only requires a one dimensional integral over $\omega_i$ rather than a three dimensional integral over $(y_i,z_i,\bz_i)$ our computations will be much simpler.  However, 4D kinematics will get in the way of a straightforward interpretation of the low point amplitudes.

Our integration can be visualized as follows.  We have a four dimensional delta function coming from momentum conservation. In a pure massless theory our amplitude will only have support when the momenta form a closed null polygon.  Meanwhile the 2D conformal correlators are labeled by the directions of the momenta.  This means that if we hold the $(z_i,\bz_i)$ fixed for each particle, we are integrating over the space of lengths we can assign to these null directions such that the polygon still closes.  If we have less than a certain number of particles, all of the Mellin integrals will be saturated by delta functions.  Moreover if there are fewer Mellin transforms being performed than number of delta functions, some constraints on the $z_i$ will remain. We note that for the 4D case this number is actually five since once we have a closed null polygon, any overall dilation of it will again be a closed null polygon.  Indeed, we can break up our multi-Mellin transform into a simplex integral and an overall scale
\begin{align}\label{eq:simplex}
\prod_{i=1}^n  \int_0^\infty d\omega_i  \, \omega_i^{i\lambda_i } [...]= \int_0^\infty ds s^{n-1+i\sum\limits_i\lambda_i}\prod_{i=1}^n  \int_0^1 d\sigma_i  \, \sigma_i^{\lambda_i }\delta(\sum\limits_i\sigma_i-1) [...]\,,
\end{align}
where  $s\equiv\sum\limits_i \omega_i$ and $\sigma_i\equiv s^{-1}\omega_i\in[0,1]$  so that $\sum\limits_{i=1}^n\sigma_n=1$.  At this point, there are two things to note.  One, going to lower dimensions means fewer delta functions, which will mean fewer restrictions on the $z_i$'s.  For example this was used in~\cite{Lam:2017ofc} to make the conformal block decomposition of the now-non-singular four point function more tractable.   Second, our transform probes all energy scales.  While, on the one hand, this means that one might be out of luck if you only know the momentum space amplitude in a low energy effective theory, on the other hand once/if a dual is established, we see that the correlators will capture the UV and IR behavior of the bulk theory.  In particular~\cite{Stieberger:2018edy} was able to evaluate this transform on string amplitudes where the soft UV behavior plays an essential role in making the integral over the overall energy scale ($s$ in~(\ref{eq:simplex})) converge.  For the gluon examples we consider here, we have
\begin{align}\label{eq:sc}
\mathcal{A}_{\ell_1\cdots\ell_n} (\Lambda\omega_i ,z_i ,\bar z_i)\,=\Lambda^{-n} \mathcal{A}_{\ell_1\cdots\ell_n} (\omega_i ,z_i ,\bar z_i)\,
\end{align}
where $\mathcal{A}$ includes the momentum conserving delta function. This implies that the factor of $s^n$ in~(\ref{eq:simplex}) will cancel out leaving an integral of the form~(\ref{m02pt}) over $s$ and giving a delta function restricting the sum of the $\lambda_i$'s to be 0. This characteristic is generic to any amplitude in a theory that is conformally invariant, as tree level gluon scattering is here.  Specifically, we have
\be\label{eq:mts}
\scalemath{0.9}{\mathcal{\hat A}_{J_1\cdots J_n}(\lambda_j, z_j,\bar z_j)
=2\pi\delta(\sum\limits_i\lambda_i)  \prod_{i=1}^n  \int_0^1 d\sigma_i  \, \sigma_i^{i\lambda_i } \,
A_{\ell_1\cdots \ell_n} (\sigma_j,z_j,\bar z_j)\delta^{(4)}(\sum\limits_i\e_i\sigma_iq_i)\delta(\sum\limits_i\sigma_i-1)}\,
\ee
where $\e_i=\pm1$ for outgoing (incoming) particles and $q_i^+=2$ as in~(\ref{eq:nullref}).  

Next, we will consider 3pt and 4pt stripped MHV amplitudes.  Each of the Mellin frequency integrals will be saturated, so that the relative energies $\sigma_i$ are solved for in terms of the $(z_i,\bz_i)$.  The momentum space MHV amplitude can then just be evaluated at these $\sigma_i(z_j,\bz_j)$ and the momentum conserving delta function will provide additional constraints as well as certain Jacobian factors.  Moreover, since $\sigma_i$ in~(\ref{eq:mts}) are constrained to lie on the unit simplex, different 4D crossings (choices of $\e_i$) will have different support on the celestial sphere. 

For the 3 point function, we have to go to $(2,2)$ signature to get interesting results.  This comes from the fact that $1\rightarrow 2$ massless decays are forced to be collinear by kinematics (if not, one could go to the rest frame of the final state).  However, with another timelike direction we can find a non-degenerate closed null triangle.  In the signature $(-+-+)$ we have $z$ and $\bz$ independent real variables, and the celestial sphere is now Lorentzian.  The Lorentz group now becomes  $SL(2,\mathbb{R})\times SL(2,\mathbb{R})$ where the two factors act separately on $z$ and $\bz$.
The tree-level color-ordered MHV three-point amplitude
\begin{align}
\mathcal{A}_{--+}(\omega_i ,z_i ,\bar z_i) &= {\< 12 \>^3 \over \<23\> \<31\>}\, \delta^{(4)} (p_1^\mu + p_2^\mu+p_3^\mu )\notag\\
&=-2{\omega_1 \omega_2 \over \omega_3 }{z_{12}^3 \over z_{23} z_{31}}\, 
\delta^{(4)} (\sum_i \varepsilon_i \omega_iq_i^\mu )\,,
\end{align}
here written in terms of spinor helicity variables~\cite{Elvang:2013cua} -- which up to a little group transformation can be taken to be
\be\scalemath{1}{
|p\rangle =\e \sqrt{2\omega} \left(\begin{array}{c}- 1 \\-z\end{array}\right)\,,~~~~~~|p]  = \sqrt{2\omega} \left(\begin{array}{c}-\bar z \\ 1\end{array}\right)}\,,
\ee
in terms of $(\omega,z_i,\bz_i)$ where
$p^\mu  =\e\omega (  1+z\bar z , z+\bar z, z-\bar z, 1-z\bar z)$ in this signature
 -- gets transformed to
\begin{align}\label{--+}
\,\mathcal{ \hat A}_{--+} (\lambda_i ;z_i,\bar z_i)  
= -
{\pi} \, \delta(\sum_i\lambda_i){\mathrm{sgn}(z_{12}z_{23}z_{31})\delta (\bar z_{13} ) \delta(\bar z_{12}) 
\over |z_{12}|^{-1- i \lambda_3}  |z_{23}| ^{1- i\lambda_1} |z_{13}| ^{1-i\lambda_2}}
\,\prod_{i=1}^3
\mathbf{1}_{[0,1]}(\sigma_{*i}),~{z_i,\bz_i\in\mathbb{R}}\,
\end{align}
where $\mathbf{1}_{[0,1]}(x)$   is the 
 indicator function
 \begin{align}\label{indicator}
 \mathbf{1}_{[0,1]}(x) =
 \begin{cases}
 1\,,~~~~~\text{if}~~x\in [0,1]\,,\\
 0\,,~~~~~\text{otherwise}\,.
 \end{cases}
 \end{align}
Here we have solved the momentum conserving delta function constraints assuming $z_{ij}\neq0$ so that the MHV amplitude is non-zero.  For the anti-MHV case there is a similar locus of support in which the $\bz_{ij}$ are non-zero and $\delta (z_{13} ) \delta(z_{12}) $ appears.  Note that the support of the indicator function serves to select different ordering of $z_i\in\mathbb{R}$ depending on the scattering channel $ij\ce{<-->} k$
\be
\prod_{i=1}^3
\mathbf{1}_{[0,1]}(\sigma_{*i}):~~
\begin{array}{llll}
a)~~~ 12 \ce{<-->} 3~~~&\Rightarrow z_1<z_3<z_2  ~\mathrm{or}~ z_2<z_3<z_1&\\
b)~~~ 13 \ce{<-->} 2~~~&\Rightarrow z_1<z_2<z_3  ~\mathrm{or}~ z_3<z_2<z_1&\\
c)~~~ 23 \ce{<-->} 1~~~&\Rightarrow z_3<z_1<z_2  ~\mathrm{or}~ z_2<z_1<z_3&.\\
\end{array}
\ee
Here, both the ordering of the $z_i$ and the crossing channel change under $SL(2,\mathbb{R})\times SL(2,\mathbb{R})$ via $\varepsilon_i \to\varepsilon_i \text{sgn}((cz_i+d)(\bar c \bar z_i +\bar d))$, leaving the indicator functions invariant.  
  One can check that~(\ref{--+}) indeed transforms like a three point function of spin-one primaries with weights
\begin{align}
\begin{split}
&h_1 = {i\over 2} \lambda_1 \,,~~~~~~~~~~~\bar h_1 = 1+ {i\over 2} \lambda_1\,,\\
&h_2 =  {i\over 2} \lambda_2 \,,~~~~~~~~~~~\bar h_2 = 1+ {i\over 2} \lambda_2\,,\\
&h_3 = 1+ {i\over2} \lambda_3 \,,~~~~~~\bar h_3  = {i\over 2}\lambda_3\,.
\end{split}
\end{align}
The corresponding expressions for the anti-MHV amplitude can be found in~\cite{Pasterski:2017ylz}.

We can now return to $(-+++)$ signature to evaluate the transformed $4$-pt tree-level color-ordered MHV amplitude.  Starting from 
\begin{align}
\mathcal{A} _{--++}(\omega_i,z_i,\bar z_i)= {\langle 12\rangle ^3 \over \langle 23\rangle \langle 34\rangle \langle 41\rangle}
\delta^{(4)} (  \sum_{i=1}^4 \varepsilon_i  \omega_i q_i )
\,,
\end{align}
we find
\begin{align}\label{eq:4ans} 
\mathcal{\hat A} _{--++}(\lambda_i,z_i ,\bar z_i)=& -{\pi\over4}  \delta( \sum_k \lambda_k)  \delta\left({|z-\bar z|\over2}\right) \,
\left(  \prod_{i<j}^4 z_{ij}^{\frac h3 -h_i -h_j} \bar z_{ij}^{\frac {\bar h}{3} - \bar h_i -\bar h_j}\right)\,
 z^{\frac 53}\,(1-z)^{-\frac13}\, \notag\\
& ~~~\times\prod_{i=1}^4\mathbf{1}_{[0,1]}(\sigma_{*i})\, \end{align} where the weights are
 \begin{align}\label{4weight}
\begin{split}
&h_1= {i\lambda_1\over2}\,,~~~~~~~~~h_2= {i\lambda_2\over2}\,,~~~~~~~~~\,
h_3= 1+{i\lambda_3\over2}\,,~~~~h_4= 1+{i\lambda_4\over2}\,,\\
&\bar h_1= 1+ {i\lambda_1\over2}\,,~~~~\bar h_2=1+ {i\lambda_2\over2}\,,~~~~
\bar h_3= {i\lambda_3\over2}\,,~~~~~~~~~\bar h_4= {i\lambda_4\over2}\,.
\end{split}
\end{align}
Now the four massless particles are constrained to lie on a `celestial circle.' (Recall global conformal transformations of the complex plane map circles to circles and that any three points can be mapped to $\{0,1,\infty\}$.  Momentum conservation would imply that the fourth point also lie on the equator of the celestial sphere.)  Here $(z,\bz)$ are the conformal cross ratios
\begin{align}\label{crossratio}
z\equiv  {z_{12} z_{34} \over z_{13}z_{24}}\,,~~~~~~\bar z\equiv  {\bar z_{12} \bar z_{34} \over\bar z_{13}\bar z_{24}}\,,
\end{align}
so that the delta function appearing in~(\ref{eq:4ans}) restricts the cross ratio to be real.  The indicator functions then imply that different $2\rightarrow2$ channels have support for different intervals of this cross ratio
\be
\prod_{i=1}^4\mathbf{1}_{[0,1]}(\sigma_{*i}):~~\begin{array}{lll}
a)~~~ 12 \ce{<-->} 34~~~&\Rightarrow~~~ 1< z \\
b)~~~ 13 \ce{<-->} 24~~~&\Rightarrow~~~ 0<z<1\\
c)~~~ 14 \ce{<-->} 23~~~&\Rightarrow~~~  z<0\,
\end{array}
\ee
and are vanishing for other 4D crossing channels, e.g. $1\rightarrow3$.  Returning to $(-+-+)$ signature, one can verify that the 3pt MHV and anti-MHV amplitudes appropriately glue together to form the 4pt amplitude via an analog of BCFW. 

The hope would be to eventually connect such recursion relation statements on the 4D amplitudes side to an OPE statement in the CFT, but we are not there yet.  Beyond the current/primary OPE-like statements that drove our soft physics endeavor, some recent statements about an OPE-like structure from collinear limits have been made in~\cite{Fan:2019emx}.  At the moment the order of limits implied in our transform (keeping $(z_i,\bz_i)$ fixed and solving for $\omega_i$ consistent with the kinematics) precludes taking soft limits of low point amplitudes (since assumed not to be collinear).  However by looking at subsectors of higher point amplitudes one should be able to avoid such problems.  Indeed, the main subtleties we have encountered in the examples of this section have boiled down to how translation invariance is handled.  Some attempts at formalizing this in terms of shifts in the weights (related to analytically continuing off the principal series) can be found in~\cite{Stieberger:2018onx}.  Meanwhile progress has been made in evaluating these transforms for different theories, in addition to the string computation in~\cite{Stieberger:2018edy}, Mellin transforms of $n$-point $N^k$-MHV amplitudes have been worked out in~\cite{Schreiber:2017jsr}.

~\\

\section{What Lies Ahead}\label{sec:unitary}

We are at a stage where we have motivated the existence of an enhanced symmetry group, recognized that the same low energy modes that prove their relevance hint at a celestial sphere holographic dual, and finally shown that we can map the finite frequency scattering states to the 2D picture as well.  But it is reasonable to ask what we have gained thus far and also what we can expect to learn in the long run.  At first glance we have made our lives harder by performing the change of basis.  The integral transformations take simple momentum space amplitudes to functions with singular support, non-integral scaling dimensions, and less obvious OPE behavior.  However, if we do begin to build a better understanding of what CFTs would be dual to scattering in asymptotically flat spacetimes, one would potentially have much to gain by the reverse map.

It has been evident for a while that perturbation theory for gauge theory amplitudes does not do justice in capturing the simplicity of the final results.  We have seen various attempts to attack this problem, amongst them on-shell recursion relations~\cite{Britto:2004ap,Britto:2005fq}, the scattering equations~\cite{Cachazo:2013gna,Cachazo:2013hca,Cachazo:2013iea, Cachazo:2014xea}, 
the twistor and ambitwistor string~\cite{Penrose:1967wn,Witten:2003nn,Mason:2013sva}, and the amplituhedron~\cite{Arkani-Hamed:2013jha}.  Our efforts would be adding one more to the list if the dual perspective could be shown to shed light.  At the same time, it seems reasonable that what we are up to is not completely unrelated to the above efforts.  Such connections should be fleshed out.

On a practical side, there are various routes to move forward along.  For one, our understanding of the map from 4D to 2D has been rather example driven and focused on position space wave functions.  By looking instead at how the states are constructed in terms of the 4D Hilbert space and Poincar\'e generators, it seems that one can make more general statements: for instance, reinterpreting the results of~\cite{Stieberger:2018onx} on the effect of translations; or the manner in which 4D unitarity manifests itself on the 2D side. Some statements about the optical theorem have been made in~\cite{Lam:2017ofc}.  It would be reasonable to expect to gain a more informed understanding of the manner in which Hermitian conjugation, combined with a parity transformation, are needed to get the 4D operators to map to 2D operators with the standard Hermiticity conditions there -- i.e. we really need to understand the representations we are dealing with better.  The principal series was used for its completeness relations and the fact that the states have finite inner products.  They also appear in the CFT bootstrap literature, but as an analytical tool.  

 Moreover can we find examples~\cite{Gadde:2017sjg} of actual known 2D CFTs that are in the image of this map?  While our momentum space transforms are reminiscent of perturbative AdS computations, we have no locality in momentum space except for in the contrived example of a near-extremal decay.  How do we expect our 2D correlators to behave?  We can try to be more formal and understand the implications of having not just a stress tensor but also a current corresponding to the supertranslations.  Can we demonstrate an OPE-like statement, and would this have a chance of teaching us how to better handle collinear divergences in field theory~\cite{Fan:2019emx}?  Does the string worldsheet CFT connect to the celestial sphere CFT~\cite{Fan:2019emx}?  How do our results connect to AdS/CFT proper, either in a small curvature limit or from the point of view of hyperbolic foliations~\cite{0303006dBS}?  We are not lacking in questions we should be trying to answer, and it's time to stop writing and start computing...

\pagebreak

\pagebreak\addcontentsline{toc}{section}{References}
\begin{spacing}{1}
\bibliography{ArXivThesis}
\bibliographystyle{utphys} 
\end{spacing}

\end{document}